# 过冷液体铝成核：基于无轨道密度泛函理论的第一性原理分子动力学研究

白子豪

**摘要**：成核是非常常见的发生在微观尺度的物理现象。但由于成核势垒的存在，成核过程的时间尺度通常远远高于微观粒子动力学的时间尺度。这些特点使得成核过程的研究无论是在实验、理论还是计算机模拟角度都非常具有挑战性。在本工作中作者探讨了将基于无轨道密度泛函理论（OFDFT）的第一性原理计算应用于过冷液体成核模拟的意义与可行性。为分析 OFDFT 在过冷液体成核模拟中的表现，作者采用无轨道密度泛函理论分子动力学（OFDFT-MD）计算了铝的熔点并模拟了过冷液体铝的成核过程。通过使用作者开发的成核问题序参量计算软件 GlassViewer 对成核过程中的键朝向序参量（BOO）演化与平均首次通过时间（MFPT）进行计算，从而对成核过程中的原子局域结构变化以及前驱体的成核动力学参量进行分析，并与使用嵌入原子势方法（EAM）半经验势能计算的结果进行对比分析。结果显示 OFDFT 可以给出相对可靠的熔点，其预测的成核参数也与 EAM 势能所计算结果相符合。本工作显示了使用 OFDFT 进行过冷液体成核模拟的可行性，这使得利用这一第一性原理方法探讨电子结构以及电磁场在过冷液体成核过程中的影响成为可能。

**关键词**：成核；无轨道密度泛函理论；过冷液体铝；分子动力学

**Abstract:** Nucleation is very common physical phenomena that occur at the microscopic scale. However, due to the presence of nucleation barriers, the time scale of nucleation is usually much longer than that of microscopic particle dynamics. These characteristics make it significantly challenging to study nucleation experimentally, theoretically, or computationally. In this work, the authors explore the significance and feasibility of applying first-principles calculations based on orbital-free density functional theory (OFDFT) to simulate nucleation in supercooled liquids. To analyze the performance of OFDFT in simulating nucleation in supercooled liquids, the authors employ orbital-free density functional theory molecular dynamics (OFDFT-MD) to compute the melting point of aluminum and simulate nucleation in supercooled liquid aluminum. By utilizing GlassViewer, an order parameter calculation software for nucleation problems developed by the authors, the evolution of the bond-orientational order (BOO) parameters and the mean first-passage time (MFPT) are calculated to analyze the changes of local atomic structures during the nucleation processes and the nucleation kinetics parameters of precursors. The results are then compared with those obtained using the embedded atom method (EAM) semi-empirical potential. The results indicate that OFDFT can provide relatively reliable melting points, and the predicted nucleation parameters are consistent with the results obtained using the EAM potential. This work demonstrates the feasibility of employing OFDFT for simulating nucleation in supercooled liquids, enabling the exploration of the effects of electronic structures and electromagnetic fields on nucleation processes in supercooled liquids using this first-principles approach.

**Keywords: Nucleation; Orbital-Free Density Functional Theory; Supercooled Liquid Aluminum; Molecular Dynamics**

# 目录

# 第1章绪论

下面我们分别简要介绍：一阶相变与成核现象的性质、成核研究的产业价值、成核的研究困难、成核模拟的一些研究缺口与本工作的简介与意义。

在现代相变分类中，往往将相变分为两类：一阶相变与连续相变。其中一阶相变在相变点处的序参量随热力学状态参量的变化而发生突变；而连续相变在相变点处序参量随热力学状态参量的变化连续变化。与连续相变不同，一阶相变在其热循环中可能呈现“迟滞性”（hysteresis），即系统在相变点外一定范围一段时间内仍保持在原相，而未产生相变的现象，此时原相处于亚稳相，由于热力学势垒的存在，无法向稳态演化。一阶相变中的迟滞现象极大的增加了一阶相变的研究难度，并引发出了丰富多彩的物理现象与研究领域，如亚铁磁材料在低温时被施加改变方向的外磁场时可以呈现磁迟滞性；液体在快速过冷时未结晶可以形成“玻璃体系”。关于玻璃体系的理论研究一直为凝聚态领域中相对困难的问题，Science 杂志在 2005 年刊发的 100 个科学问题中便对“玻璃态是如何形成”提出理论质问[1]；新奇的玻璃体系亦亟待研究，如金属体系形成玻璃后则会展现出丰富的新奇特性。

在温度相变中，一阶相变的迟滞性表现为过热现象与过冷现象，并常在相变时伴随以“成核现象”（nucleation），即系统首先由于涨落偶然形成一个微小的新相的“核”，而后系统在核的表面快速生长为新的宏观相。在下文中，我们均只讨论液体过冷相变中的成核过程。 成核现象在生活中最常见的例子莫过于过冷水的成核：将纯净液水装在表面光滑的容器内，在温度低于熔点一定温度范围内，系统可较长时间处于液态而无法固化。直到水分子在器壁表面、或在液体内部涨落聚集为微小冰晶，而后微小冰晶迅速生长为宏观冰晶体。其中成核过程可根据成核位置分为两类：在第三相（如器壁、杂质颗粒）表面上成核的异质成核，以及在液体内部形成的均质成核，均质成核的概率远远低于异质成核。需要尤其注意的是：1.如地震、经济危机、通讯网络崩溃等，成核现象是一类由涨落驱动的低概率的“稀有事件”（rare event）[2]，作为微观事件的成核所发生的时间尺度要远远高于微观粒子动力学的时间尺度。其具体发生时间不可预测，但可给出概率表述：即单位体积与单位时间内平均成核的数量，即成核速率。2.成核速率随过冷度呈现百数量级的差异。一组数据可给人以非常的直观感受：如图 1 所中关于水的均质成核的理论计算与实验数据表明，过冷度为

35K 时，水的成核速率约在 $10^{11}m^{-3}s^{-1}$ 量级，在一杯 500ml 的纯净水中，1s 可以形成约 $10^8$ 个冰核；而过冷度约为 20K 至 25K 时，具有地球水圈体积的水在宇宙年龄的时间尺度内才能形成一个均质冰核。这意味着在过冷度低于 20K 时，我们星球上的水只能通过异质成核在第三相表面形成[3]。这个讨论一方面展现了成核速率在不同条件下的天壤之别；另一方面侧面暗示了成核研究领域中实际存在的巨大困难：不同理论模型、计算模拟与实验数据之间往往具有数量级的差异。

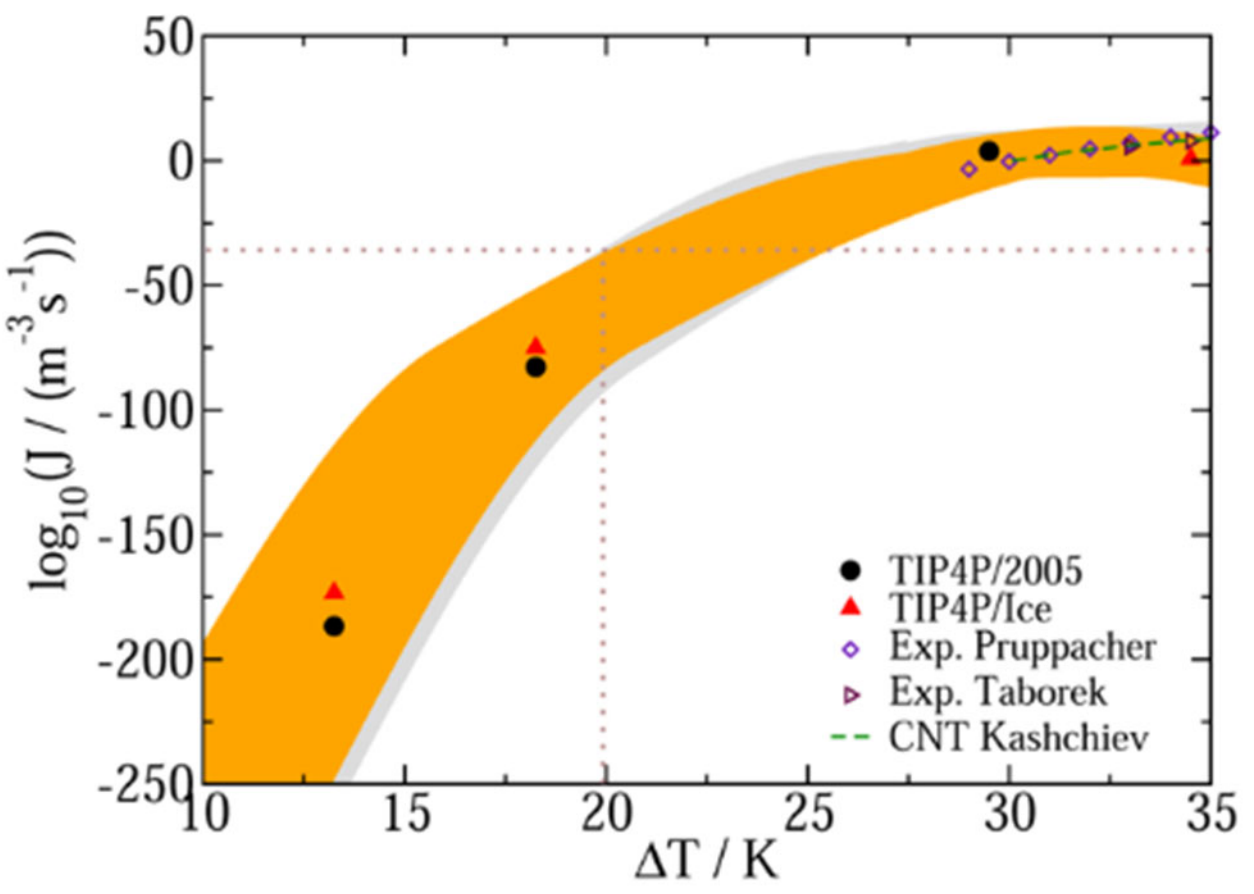

图 1：水的成核速率（对数化）关于过冷度的关系。实心黑色与红色标记代表被引用文章的理论计算结果；空心紫色红色标记代表其他文章的实验数据；绿色虚线为其他文章的经典成核理论预测结果；灰色与橙色阴影分别代表被引用文章对 TIP4P/2005 与 TIP4P/Ice 两组理论模型预测结果所估计的误差（经过插值）。水平虚线代表宇宙年龄下在地球水圈形成一个核的成核率；纵向虚线代表“低于此过冷度，在地球上水的均质成核是不可能的”。图片摘取自参考文献[3]。

对成核的研究在冶金、制药、气象等不同领域发挥着巨大的价值。不妨以其在有机合成中发挥的巨大价值作为例子：在制药等领域中有机溶液成核方式、成核品质往往决定了药物等产物的品质，但成核在分子体系很难发生并且可能伴随着多种方式：1. 分子体系相互作用力弱（与冶金中的金属间相比），往往需要数天、数周才能成核结晶，但是为了增强结晶品质，过饱和度不能太高。因此在低饱和度溶液中增强成核速率的方式亟待探索；2. 有机分子在不同的条件下往往可以形成不同结构的核，并结晶至不同的亚稳态，这被称为成核中的“多相选择（polymorphism）”问题。因此如何对成核结晶过程进行精准的控制，成为很重要的问题。近年不同强度、极化方式的激光被发现可以控制成核率与成核的多相选择。激光调控成核也在被广泛的研究。

尽管成核现象在生活中非常常见，且具备重要的产业价值，但成核的研究一直非常困难。对于实验来说，从统计的角度，实验上可以研究的过饱和度范围非常局限，过高和过低的饱和度都导致成核出现的时间过快或者过慢；从微观机制的角度，成核过程往往尺寸太小且发生的太快了，据作者所知，仅在胶体等粒子较大的体系，对成核过程中粒子的构型进行跟踪测量是可能的[4]。对于理论模型来说，以经典成核理论为主的类似的唯象模型受限于过于粗糙的毛细近似（Capillary approximation），把微观尺度的核团簇当作宏观物体进行近似处理。在经典成核理论中，认为核团簇是球形的，成核过程中仅存在两个相，微观相的化学势与宏观相相同，微观相的表面张力与平面表面张力相同。这种近似显然过于粗糙，并造成理论的失效（尤其对过饱和度很高、临界核很小的情况）。实际的成核过程内，系统可能经历两个及更多相的演变；微观团簇的性质不能用宏观相的属性进行直接计算；微观的核也不是均匀球型的。理论与实验的艰难进展促使计算机模拟成为成核研究中重要的工具。然而由于成核（在微观的意义下）“稀有事件”的本质，基于微观的相对准确的计算模型难以对如此小的概率进行抽样，亦或对如此长的时间进行模拟。其可模拟的过冷度（或降温速率）往往极高，导致模拟中的成核速率（与实验相比）极大。在（半）经验势能的模拟工作[5]中，约$10^6$个铁原子在 200ps 时出现约 35000 个核，我们估算其成核速率约为$3\times10^{38}m^{-3}s^{-1}$；图 1 所展示的成核速率的两组实验结果则约为$10^{12}m^{-3}s^{-1}$。其悬殊的数量级差异意味着若想将关于成核速率的计算结果与实验进行对比，无论对实验技术亦或模拟技术而言，挑战都是过大的。

既然难以直接对成核问题中非常关心的成核速率进行有比对意义的直接计算，计算机模拟往往可以发挥其“微观尺度”的特性，对成核过程中原子微观结构的变化进行揭秘。然而我们必须承认相变问题中涉及的原子环境是复杂的，需要较为精准的微观模型以对微观势能面进行合理建模，以得到相对精准的有效结果。然而即便在如此高的降温速率下，成核“稀有事件”的本质也使得模拟成本相当高。根据本工作中进行的尺寸依赖性测试，过冷液体铝的成核模拟中体系尺寸需达到 864 原子才可在本工作的模拟条件下避免尺寸效应的出现；同时本工作的模拟结果显示，即便对于很高的降温速率（10K/ps），体系依然需要 10ps 至 100ps 的时间以产生成核。计算化学领域公认的具有较高精度的第一性原理计算方法为 KS 密度泛函理论(KSDFT)，一个使用 KSDFT 对过冷液体铌成核的模拟中使用的模拟尺度为 128 原子与 18ps[6]，由于

KSDFT 的计算量为非线性标度，极难将其扩展到更大的空间时间尺度。另一方面，对于（半）经验势能而言，其过于简单的近似模型不一定能预测出体系成核过程中真正的微观结构演化。有研究表示，即便是两个对平衡态固体与液体结构预测的非常相似的（半）经验势能，也可能在快速降温的远离平衡的成核过程中表现出非常不同的成核动力学特征[7]。

因此亟待一种相对合理的微观模型，既基于第一性出发，以对势能面上不同结构的原子环境提供第一性的描述，同时也在 KSDFT 的基础上做合理近似，以一定的精度为代价，换取对于成核所需时间尺度、空间尺度可行的计算能力。无轨道密度泛函理论（Orbital-Free Density Functional Theory，OFDFT）恰好满足这样的特性。其基于第一性原理，但具有线性标度的时间复杂度[8]。OFDFT 曾被应用于合金的位错成核研究中[9,10]，但据作者所知，还未有将其应用于过冷液体的成核的研究。在本工作中作者首先进行 OFDFT 框架下的熔点测试，而后实施 OFDFT 分子动力学框架下的金属铝的过冷液体成核模拟，并对成核过程中的原子局域结构的变化与前驱体的成核参数进行分析。同时，作者亦使用一个在熔点附近拟合的 EAM 半经验势能进行相同的模拟，并将两个计算框架下的结果进行比较。结果显示，OFDFT 所模拟的熔点在 12.15%的误差内与实验结果一致；成核过程中的原子结构变化与前驱体的成核动力学参数亦与在熔点附近拟合的 EAM 半经验势能给出的结果相符合。我的工作的结果揭示了使用 OFDFT 分子动力学进行过冷液体成核研究的可能性。由于第一性原理模拟引入了电子自由度，我的结果亦为过冷液体成核过程的电子结构研究与外场调控模拟提供了可能性。

# 第2章 理论与方法

## 2.1 经典成核理论

经典成核理论（Classical Nucleation Theory，CNT）创建于 20 世纪二十至四十年代 [11–13]，尽管过于简化的 CNT 与实验相比被发现有较大偏差，作为最经典的成核理论，CNT 为成核过程提供了非常直观的基础框架；在这里，我们通过对 CNT 的介绍，描绘成核的基本图景。

为了对 CNT 提供更好的理解，我们从 1982 年 Kashchiev 在其论文中使用的一般形式出发[14]，对于以团簇粒子数为反应坐标的成核过程，在原相中形成一个核团簇自由能的增量应满足下式：

$$\Delta G(n, \Delta\mu) = -n\Delta\mu + F_s(n, \Delta\mu) \tag{2.1-1}$$

其中$n$为团簇的粒子数，$\Delta\mu$为平均每粒子的新旧相的化学势差，等式右侧第一项为不考虑界面效应的情况下，仅考虑体相变化产生的自由能增强；实际的自由能增量并不能完全由第一项表示，将额外的自由能增量记为$F_s(n, \Delta\mu)$，这个额外的自由能增量是由于界面附近物质的结构与性质与体相有所区别所引起的。

不同的理论对后一项有不同的处理，经典成核理论使用所谓的“毛细近似”，将微观层面的团簇视为宏观物体进行处理。当应用至液固相变时，其至少包括以下近似[15]：

1. 团簇被视为具有确定半径的均匀同质的球体，其内部具有液体的体相性质，外部具有气体的体相性质。（这意味着只考虑各向同性）
2. 团簇的表面能被近似为$\gamma_\infty(T)$与$A(n)$的乘积，其中$\gamma_\infty(T)$为温度$T$下的平界面表面张力，$A(n)$为团簇的表面积。（注意：1.这一近似非常强，实际的表面张力依赖于界面曲率，尤其对于微观尺度的小团簇；2.这里仅写出表面张力对温度的依赖，未写出对压力等其他的依赖。）

这意味着从原相中出现粒子数为 n 的团簇的自由能增量为:

$$\Delta G(n, \Delta\mu) = -n\Delta\mu + \gamma_\infty A(n) \quad (2.1\text{-}2)$$

$$A(n) = (36\pi)^{1/3}\rho_s^{-2/3}n^{2/3} \quad (2.1\text{-}3)$$

其中$\rho_s$为新相的体相粒子数密度。

这意味着在成核的过程中，出现如图 2 所示的自由能的成核势垒，跨越势垒是典型的激活过程（activated process），这意味着系统在相变前需要首先通过随机涨落形成一个跨越势垒的团簇，这解释了为何系统的成核相变需伴随以随机的初始等待时间。

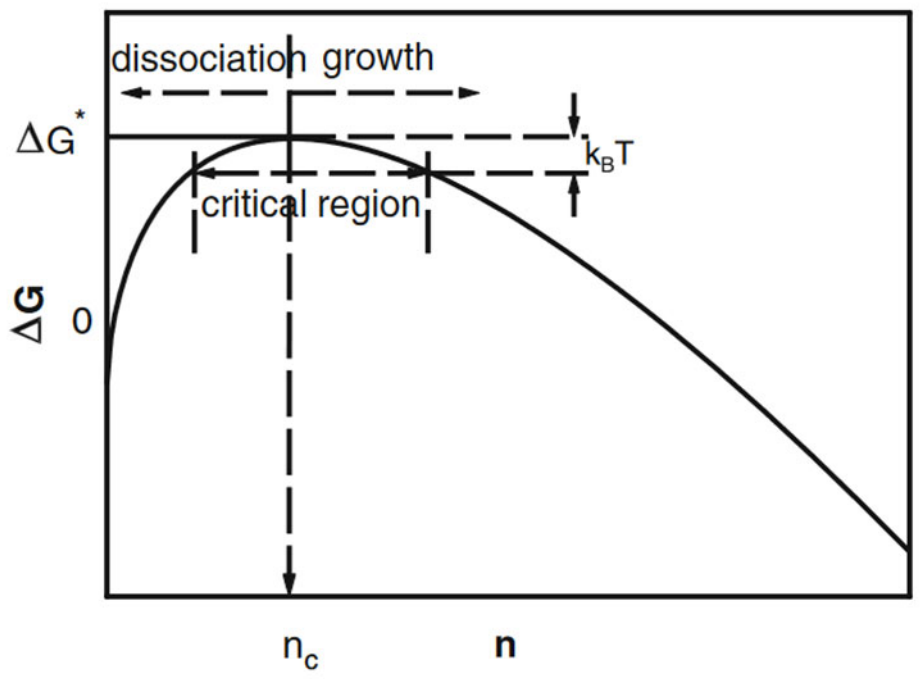

图 2：成核过程中自由能关于团簇原子数 n 的曲线。图片摘选自参考文献[15]。

自由能差值曲线极大值处被称为自由能势垒（即成核势垒，nucleation barrier），其对应的粒子数被记作$n_c$，具有$n_c$粒子数的团簇被称作临界团簇（critical cluster）

$$n_c = \frac{32\pi}{3}\frac{\gamma_\infty^3}{\rho_s^2\Delta\mu^3} \quad (2.1\text{-}4)$$

$$\Delta G^* = \frac{16\pi}{3}\frac{\gamma_\infty^3}{\rho_s^2\Delta\mu^2} \quad (2.1\text{-}5)$$

经典成核常用于预测成核速率，略去推导，CNT 经过马尔可夫过程近似等更多近似后计算的稳态成核速率为[15]：

$$J = J_0 exp\left(-\frac{\Delta G^*}{k_B T}\right) \quad (2.1\text{-}6)$$

其中$J_0$为动力学前因子（Kinetic Prefactor），其写作如下形式

$$J_0 = \mathcal{L}(\nu A^*)\rho_1 \quad (2.1\text{-}7)$$

$$\mathcal{L} = \sqrt{-\frac{\Delta G''(n_c)}{2\pi k_B T}} \qquad (2.1\text{-}8)$$

$\mathcal{L}$被称作 Zeldovich 因子,其反映了处于自由能极大值的临界核团簇跨越势垒变大（而非返回变小）的概率；$\nu A^*$是临界核团簇被一个新粒子所依附的速率（注意其仅为“正向速率”，需考虑团簇损失粒子的“反向速率”后才可得“净速率”），其与具体相变种类中的微观机制有关；$\rho_1$是相变前的相的粒子数密度，也反映了成核位点（在均相成核中，每一个粒子都可能是成核位点）[15]。CNT 得到的成核速率形式最早被用于过饱和蒸汽中的液体成核[13]，在凝聚体系下的成核速率也有所研究[16]。作为最经典的成核图像，在本世纪的关于蛋白质溶液与胶体悬浮液等体系的成核文献中也能看到 CNT 成核速率公式的影子[17]。

可以看到成和速率显式的随温度指数变化，这为成和速率随着温度以数量级的规模变化提供了解释，当然指数中成核势垒中的表面张力与相变化学势差值亦依赖于温度。对于动力学前因子$J_0$，有学者在研究中认为其随温度缓慢变化[18]。

须指出，对于过冷液体的成核而言。过低的过冷度与过高的过冷度都不利于成核。这是由于由式（2.1-5）与（2.1-6）可见，当温度为熔点时，两相化学势差值为 0，成核势垒无穷大，成核率为 0。同时当温度趋于 0 时，成核率中的指数项也为无穷大，成核率亦为 0。所以当温度处于 0 与熔点之间时存在成核率的极大值，过低与过高的过冷度都不利于过冷液体的成核。

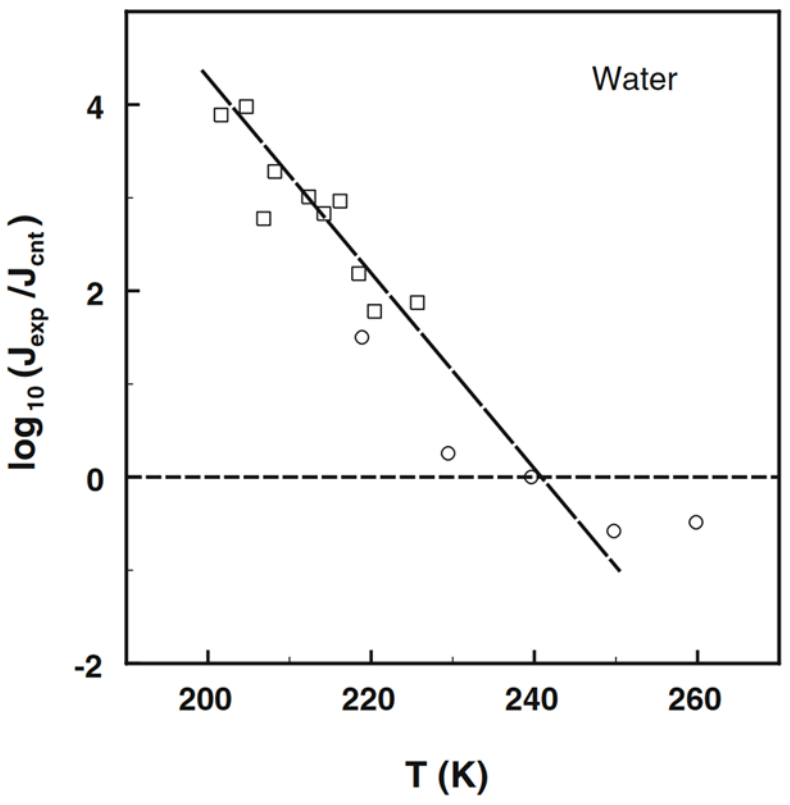

图 3：CNT 预测的水的气液成核速率与实验结果对比图。可见 CNT 预测的水的气液成核速率与实验结果具有数量级偏差，其中圆圈[19]与方块[20]离散标记为实验结果；水平虚线为 CNT 预测值，图片改编自参考文献[15]。

必须指出，CNT 是一个非常简化的理论，其仅使用粒子数作为反应坐标，无法从

多个反应坐标对“多步成核”等现象进行描述；其计算的反应速率被发现以数量级偏差于实验[15]（如图 3 为例）；其计算的自由能亦被发现与计算模拟有显著偏差[18]。随着成核研究的进行，出现较多对于 CNT 的各个方面的修正研究工作。

## 2.2 计算机模拟方法

原则上我们可以通过计算电子与原子核的运动对不同凝聚态体系的不同性质进行计算。对于凝聚态系统，电子哈密顿可写为

$$\hat{H} = -\sum_{i}\frac{\hbar^2}{2m_e}\nabla_i^2 + \frac{1}{2}\sum_{i\neq j}\frac{e^2}{|\vec{r_i}-\vec{r_j}|} + \sum_{i,l}\frac{Z_l e^2}{|\vec{r_i}-\vec{R_l}|} - \sum_{l}\frac{\hbar^2}{2M_l}\nabla_l^2 + \frac{1}{2}\sum_{I\neq J}\frac{Z_I Z_J e^2}{|\vec{R_I}-\vec{R_J}|} \qquad (2.2\text{-}1)$$

其中$\vec{r}$与$\vec{R}$分别对应电子与原子核的坐标，$m_e$与$M$分别对应电子与原子核的质量。等式右侧从左到右分别为电子动能，电子-电子相互作用能、电子-原子核相互作用能、原子核动能、原子核-原子核相互作用能。

对于较重元素的原子核可采取绝热近似（Born-Oppenheimer），将原子核和电子自由度分离：以量子力学求解电子在静止原子核构成的静止势场中的运动；对于原子核的运动单独通过经典力学的方法求解。对于仅关心原子核运动的问题，我们并不关心电子实际的运动方式，仅仅希望通过对电子体系能量的求解，得到原子（核）的势能与受力情况，进而求解原子的运动方程。对于精度允许的情况，我们也可以越过电子体系能量的计算，通过对原子间相互作用方式进行建模并以实验结果与第一性原理计算结果对参数进行拟合，而得到原子间的“力场模型”，其可以对给定的原子的坐标非常快速的计算出各原子的受力情况。当然通过量子力学求解电子能量进而给出原子受力的方法是符合“第一性原理”的，其精度通常要高于对实验与第一性原理结果拟合而得到的 “（半）经验模型”（完全基于实验数据的模型被称为经验原子间势能；部分基于实验数据，部分基于理论计算数据的模型被称为半经验原子间势能），前者的计算量也要远远高于后者。

对于给定原子核坐标求解电子体系能量的问题一般使用密度泛函理论进行计算，

我们将在 2.2.1 小节中对密度泛函理论与本研究中使用的无轨道密度泛函理论进行介绍；在 2.2.2 小节中我们将对一类本研究中使用的对金属体系近似较好的力场模型“嵌入原子法”模型进行介绍；给定原子间势能求解不同系综下原子运动轨迹的方法即分子动力学方法（Molecular Dynamics，MD），将在 2.2.3 小节中介绍。注意，在本篇论文上下文的讨论仅限于不考虑电子激发态的电子基态计算模拟。

### 2.2.1 密度泛函理论与无轨道密度泛函理论

对于给定的原子核坐标，电子哈密顿仅包含（2.2-1）式的前三项，即电子动能、电子-电子相互作用能与电子-原子核相互作用能：

$$\hat{H} = \hat{T} + \hat{V}_{e-e} + \hat{V}_{ext} \tag{2.2-2}$$

对于第一性的电子体系能量计算，由于多电子体系自由度过高，直接使用薛定谔方程由（2.2-2）中的哈密顿量求解电子波函数具有无法接受的计算成本。常通过密度泛函理论（Density Functional Theory）进行近似求解。密度泛函理论的可用性基于 Hohenberg-Kohn 定理[21]。对于我们关注的多电子体系，此定理可总结为：

1. 电子基态波函数与外势场分别可以由电子基态密度唯一确定，即存在确定泛函$\Psi[\rho]$与$V_{ext}[\rho]$。
2. 确定外势场，总能密度泛函$E_{V_{ext}}[\rho] =< \Psi[\rho]\left|\hat{T} + \hat{V}_{e-e} + V_{ext}\right|\Psi[\rho] >$取值等于体系基态能量，当且仅当$\rho$等于基态电子密度$E_0$。否则$E_{V_{ext}}[\rho] > E_0$。
3. $F[\rho] =< \Psi[\rho]\left|\hat{T} + \hat{V}_{e-e}\right|\Psi[\rho] >$与外势场无关，因此总能泛函又可写作$E[\rho] = F[\rho] + \int \rho(\vec{r})V_{ext}(\vec{r})d^3 r$。

由此，如能构造出合适的泛函$F[\rho]$，就可以对以下泛函通过自洽或优化的方式求解电子基态密度，进而得到体系基态能量。这样我们可以越过求解 3N 维的电子波函数（N 为电子数），而仅求解 3 维的电子密度，大大减少了计算成本。

Kohn-Sham 进一步为密度泛函理论提供了一套可行的计算框架[22]。其将总能泛函写为：

$$E[\rho] = \sum_i \Psi^* \left(-\frac{1}{2}\nabla^2\right)\Psi + E_{hartree}[\rho] + E_{XC}[\rho] + \int \rho(r)V_{ext}(r)dr \tag{2.2-3}$$

$$E_H[\rho]=\frac{1}{2}\int_{\Omega}\frac{\rho(r)\rho(r')}{|r-r'|}drdr' \tag{2.2-4}$$

$$\left[-\frac{1}{2}\nabla^2+V_{eff}[\rho]\right]\Psi_i=\epsilon_i\Psi_i \tag{2.2-5}$$

$$\rho=\sum_{i}^{occ}f_i\,\Psi_i^*\Psi_i \tag{2.2-6}$$

$$V_{eff}[\rho]=V_H[\rho]+V_{XC}[\rho]+V_{ext} \tag{2.2-7}$$

$\Psi_i$为引入的辅助性的单电子轨道，$f_i$为辅助单电子轨道的电子占据数（费米-狄拉克分布），Kohn-Sham 的思路在于通过引入辅助波函数，从多体动能中提取出主要部分：辅助波函数的单体动能和；从电子-电子相互作用能中提取处主要部分：电子密度的经典库伦相互作用能（Hartree 能）。并将多体动能与电子-电子相互作用中的其他部分用交换关联能表示（$E_{xc}$，与多体动能相比是小量）。其中辅助波函数可以通过有效势求解，有效势又进一步通过辅助波函数构成的电子密度更新，如此便可通过自洽的方式准确求解基态电子密度与基态电子能量。在实际操作中还需考虑电子数守恒等条件。对于$E_{xc}$则有 LDA、GGA 等不同的近似处理方式。

KSDFT 中由于需要引入辅助波函数，并且需要对波函数进行正交归一化等原因，计算成本较高，一般仅限于对百原子量级进行静态计算或简单的动态计算，难以处理成核问题这种需要较大规模、较长时间进行模拟的问题。无轨道密度泛函理论（OFDFT）通过将上述辅助波函数的单体动能和部分也通过密度泛函的方式近似给出，避免了引入辅助波函数，实现了完全的“密度泛函”，大大降低了计算成本，具有线性标度的时间复杂度[8]。然而动能密度泛函对应的能量在总能中占比较大，这一近似可能会对总能计算引入较大误差，因此 OFDFT 相比 KSDFT 而言在降低了计算成本的同时也降低了一定的精度。如何构造准确的动能密度泛函是领域的难点。

### 2.2.2 嵌入原子势方法

LJ 势能通过考虑二体相互作用，很好的对稀有气体原子间相互作用进行建模。然而对于金属体系，仅采用二体相互作用无法对相互作用很好的建模，这是由于金属体系中存在弥散在空间中的自由电子，金属原子间存在多体相互作用。为了对金属原子间相互作用进行近似建模，Daw 与 Baskes 等人提出一类嵌入原子势方法（Embeded atom method，EAM）[23,24]。对于金属体系的每个原子，其给出其势能如下：

$$E_i = F_\alpha\left(\sum_{j\neq i}\rho_\beta(r_{ij})\right) + \frac{1}{2}\sum_{j\neq i}\phi_{\alpha\beta}(r_{ij}) \tag{2.2-8}$$

其中α、β为原子$i$、$j$对应的原子种类，$F_\alpha$被称作嵌入势（embeding energy），是一种多体势，原子$i$处的嵌入势依赖于其他原子$j$在原子$i$处产生的电子密度ρ的求和。原子$j$在空间某处产生的电子密度是确定的，其仅仅依赖空间位置与原子$j$原子核间的距离长度。ϕ为金属原子之间的二体势。从形式可见，EAM 势中的多体部分非常直观的体现了一种各向同性的、由电子密度所产生的相互作用。

进一步的，Finnis 与 Sinclair 对上述模型进一步扩展[25]，考虑原子$j$在原子$i$处产生的电子密度同时依赖于原子$j$与原子$i$的原子种类，其扩展形式如下，被称为 FS-EAM 势能：

$$E_i = F_\alpha\left(\sum_{j\neq i}\rho_{\alpha\beta}(r_{ij})\right) + \frac{1}{2}\sum_{j\neq i}\phi_{\alpha\beta}(r_{ij}) \tag{2.2-9}$$

对于本课题，作者正是通过无轨道密度泛函分子动力学实现金属铝的第一性成核模拟，同时也采用铝的 EAM 半经验势能进行分子动力学模拟，并将两组结果进行对比。

### 2.2.3 分子动力学

通过对原子的动力学轨迹进行模拟，以计算体系的热力学或动力学的方法被称为分子动力学。分子动力学通常采用牛顿运动方程，以给定的原子间势能对原子的运动进行模拟。在分子动力学中常使用周期性边界条件对系统进行模拟。给定能量与体积的 NVE 系综的模拟，其计算方法是显而易见的：给定原子势能，即可通过势能的梯度得到原子受力，进而积分计算原子的运动轨迹。为了考虑在恒温或者恒压条件下对体系进行模拟，研究人员则发展了不同的“恒压器”（barostat）与“恒温器”（thermostat）技术。我们必须注意，恒压器与恒温器不能只限于恒压与恒温的功能，其至少还应具备“满足相应的系综分布”等性质。

对于压力的调控，典型的算法为 Parrinello-Rahman 于 1980 年提出 Parrinello-Rahman 方法[26,27]。这一方法将晶胞基矢作为额外的动力学变量对系统的哈密顿量进行扩展，使得在体系演化时晶胞可以变化，体系的内压可以与外压平衡，并以 3：n(n

为体系粒子数)的精度近似满足 NPH 系综分布[27,28]。对于恒温算法，一类较严格的典型算法为 Nosé-Hoover 算法[29,30]，其思想亦在于引入独立的动力学变量，对运动方程进行扩展，以使得体系自然的严格满足于正则系综分布[28]。尽管其成功的满足系综分布，但具有不具有辛结构、对简单体系不具有遍态历经性等特点[28]。

理想的分子动力学算法应尽量满足以下四个方面：准遍态历经性、时间可逆性（决定论性质的）、满足相应系综分布、满足数学上的辛几何结构[28]，在 Nosé-Hoover 理论提出后，满足更多良好性质的分子动力学理论亦被提出[28]。

但我们必须指出，除却具备唯一演化路径的 NVE 系综外，满足系综分布的分子动力学理论也不一定可以反应恒温或恒压下真实的动力学演化过程。对于不同分子动力学算法的动力学表现有相关的测试与研究[31]，对于实际而言，我们只能选择满足研究需要的尽可能“良好”的算法。

## 2.3 二步/多步成核与键朝向序参量

在进行分子动力学模拟后，我们需要对原子轨迹进行后处理（post-processing）以提取我们希望得到的物理量。对于液固成核问题（以下仅讨论原子体系，对分子体系同理），最首要的是对原子所处的相进行分辨：即判断哪些原子属于液体、属于固体、亦或属于其他中间态。必须指出，实际的成核过程绝非经典成核理论所假设的那样：过冷液体中直接涨落出基态的固体核，同时假定的边界内外的两个相具有不同的确定密度。实际模拟中出现的情况往往是非“经典的”，界面处具有一定的宽度，其中的原子结构可能是连续变化的；除此以外，最终相团簇的出现前可能先出现一个或多个处于“中间相”的团簇。这种在最终相（一般是基态）生成前，先生成一个或多个中间相（亚稳态）的成核过程被称为二步/多步成核。二步成核在较多体系中被发现[4,32]，并有相应的理论研究[33,34]，多步成核机制亦被广泛发现并研究[35,36]。二步成核中的中间相被称为“前驱体”（precursor）。如图 4 所示，在具备形成金属玻璃能力的 ZrCu 合金液体的冷却模拟中被发现首先涨落出似 FCC 的前驱体，而后在其中形成似 BCC 的晶体核团簇[37]。不同体系往往具有不同的成核机制（单步/二步/多步成核），其中中间相可能具有不同的结构特征（不同的密度序或

键朝向序等）。即便是同一体系，在不同的条件下也可能出现不同的成核机制。

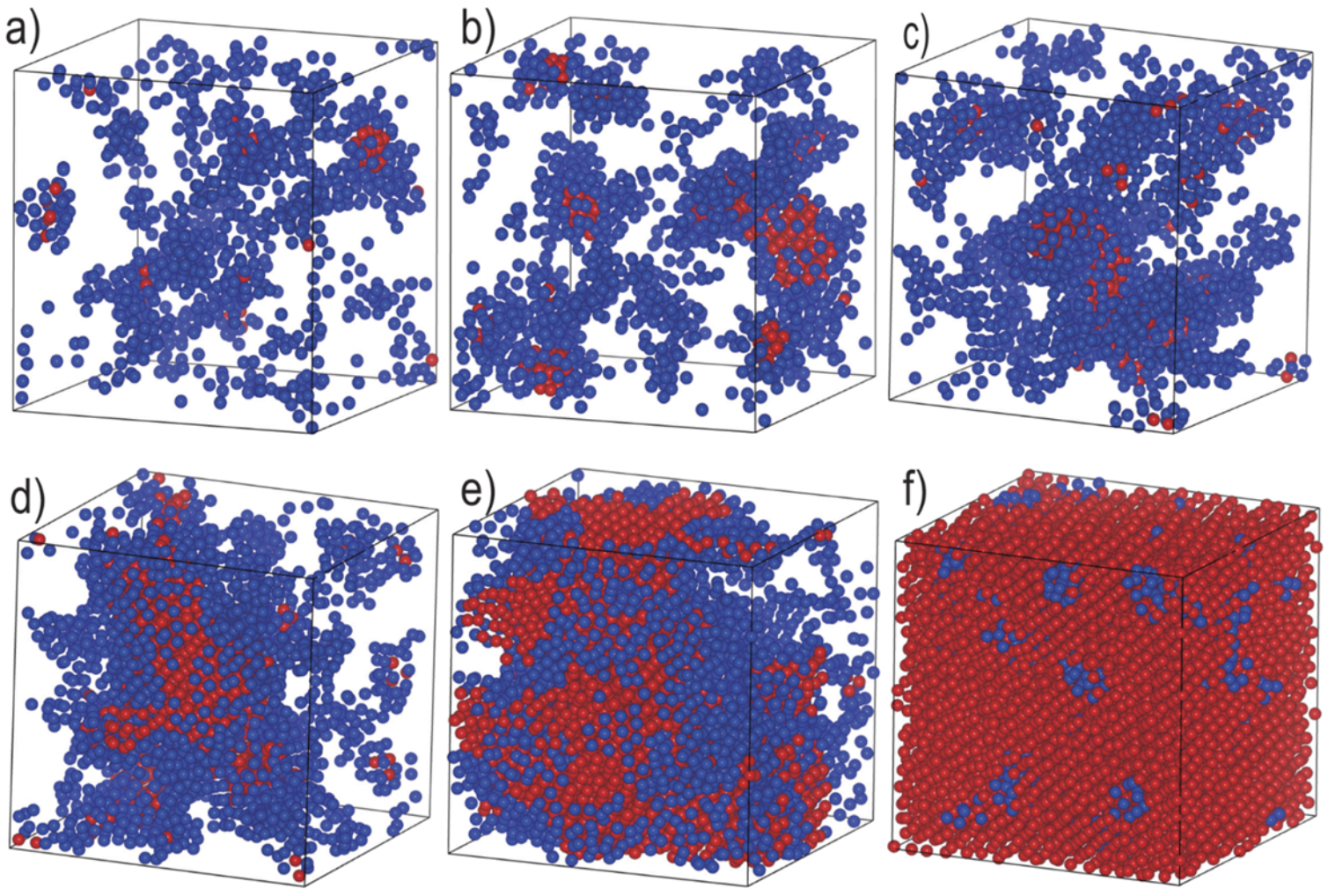

图 4：二步成核机制示例。形成金属玻璃能力的 ZrCu 合金液体的冷却模拟中被发现首先涨落出似 FCC 的前驱体，而后在其中形成似 BCC 的晶体核团簇[37]。图片摘取自参考文献[37]。

实际模拟中复杂的原子环境要求我们寻求合适的原子序参量，依照原子所处的局域环境，对不同原子所处的相进行合理的划分。液固相变中，最常见的两个原子序参量为原子的局域密度与原子的键朝向序（Bond Orientation Order，BOO）此处我们主要介绍键朝向序，通过键朝向序我们可以对原子所处的局域环境进行更加详尽的定量划分。

假设我们可以通过一定方式选取一个原子附近的最近邻原子，那么我们以这一原子为原点，称其与其最近邻原子的连线为键（bond）（这里的键仅仅代表与最近原子的连线，并非化学键），记此原子指向其紧邻原子的向量的中点为$\vec{r}$。于是通过一个原子与其各个近邻原子的键的“朝向”情况，我们可以为这一原子的局域结构提供描述。Steinhardt 等人在 1983 年发展以下键朝向序参量[38]：

$$Q_{lm}(i) \equiv \frac{1}{\mathrm{N_b(i)}} \sum_{\mathrm{j=1}}^{\mathrm{N_b(i)}} Y_{lm}\left(\vec{r}_{ij}\right) \tag{2.3-1}$$

$$Q_l(i) \equiv \left(\frac{4\pi}{2l+1} \sum_{m=-l}^{l} |Q_{lm}(i)|^2\right)^{1/2} \tag{2.3-2}$$

$$W_l(i) \equiv \sum_{\substack{\mathrm{m1,m2,m3} \\ \mathrm{m1+m2+m3=0}}} \begin{pmatrix} l & l & l \\ m1 & m2 & m3 \end{pmatrix} \times Q_{lm_1}(i) Q_{lm_2}(i) Q_{lm_3}(i) \tag{2.3-3}$$

$$\widehat{W}_l \equiv W_l / \left( \sum_{m=-l}^{l} |Q_{lm}(i)|^2 \right)^{3/2} \quad (2.3\text{-}4)$$

其中其中$i$为原子序号，$\mathrm{N_b(i)}$为原子$i$的近邻原子数，$Y_{lm}$为球谐函数，$\begin{pmatrix} l & l & l \\ m1 & m2 & m3 \end{pmatrix}$为 Wigner 3j 符号，可证明$Q_l$与$W_l$为旋转不变量，同时可见$Q_l$与$W_l$随最近邻原子数选取的更改而更改，而$\widehat{W}_l$以比值定义，一定程度可以避免由于近邻原子数选取问题造成的偏差。Steinhardt 等人在其论文中表示$\widehat{W}_l$对团簇的对称性具有较高的敏感性[38]。

除此以外，2008 年，Lechner 等人进一步发展键朝向序，通过将原子的键朝向序参量与其近邻原子的键朝向序参量进行平均，提出 averaged local bond order parameters，进一步提高了其判断晶体结构的能力[39]：

$$\bar{Q}_{lm}(i) \equiv \frac{1}{\tilde{N}_b(\mathrm{i})} \sum_{k=0}^{\tilde{N}_b(\mathrm{i})} Q_{lm}(k) \quad (2.3\text{-}5)$$

$$\bar{Q}_l(i) \equiv \left( \frac{4\pi}{2l+1} \sum_{m=-l}^{l} |\bar{Q}_{lm}(i)|^2 \right)^{1/2} \quad (2.3\text{-}6)$$

$$\bar{W}_l(i) \equiv \frac{\sum_{\substack{\mathrm{m1,m2,m3} \\ \mathrm{m1+m2+m3=0}}} \begin{pmatrix} l & l & l \\ m1 & m2 & m3 \end{pmatrix} \times \bar{Q}_{lm_1}(i) \bar{Q}_{lm_2}(i) \bar{Q}_{lm_3}(i)}{\left( \sum_{m=-l}^{l} |\bar{Q}_{lm}(i)|^2 \right)^{3/2}} \quad (2.3\text{-}7)$$

这里$\tilde{N}_b(\mathrm{i})$指的是原子$i$的近邻原子数加一（对原子$i$的近邻原子与原子$i$本身求和）。在实际中常常使用$Q_l$、$\widehat{W}_l$、$\bar{Q}_l$与$\bar{W}_l$及其他的 BOO 相关参量对体系中不同原子的局域结构的对称性进行刻画，从而判断其所处的不同的相，这里提供两种常用的做法[39]：

### 2.3.1 判断原子为液体或固体，提取成核反应坐标

取一原子与其某近邻原子的$Q_{6m}$的内积如下，其刻画了两原子之间局域结构的关联。

$$S_{ij} = \sum_{m=-6}^{6} Q_{6m}(i) Q_{6m}^*(j) \quad (2.3\text{-}8)$$

当一个键的$S_{ij}$值大于某设定阈值时（典型值为 0.5[39]），认为两原子相互“连接”。当一个原子与其近邻原子的“连接数”大于某设定阈值（典型值为 6 至 8[39]），则判定这一原子为固体；否则则认为这一原子是液态。被判定为固体。进一步的，我们按

照某种条件将不同固体原子分成团簇（比如认为相互链接的固体原子为同一团簇[39]，或间距小于一定阈值被认为属于同一团簇[40]），进而可以得到体系中当前最大的团簇的原子数$\hat{N}$，$\hat{N}$常用于作为成核过程的反应坐标。

### 2.3.2 判断原子所处的晶相类型

由于$Q_l$、$\widehat{W}_l$、$\bar{Q}_l$与$\bar{W}_l$可以很好的刻画不同原子的局域结构对称性，常通过其数值来为一个原子判断其晶相类型。在实际中常常绘制这些序参量关于反应坐标（如团簇粒子数）或反应时间的分布图像，或这些序参量之间的二维分布图像。通过这些分布，我们可以更好的对原子的相进行区分、亦或研究体系的不同相随时间的演化情况。图 5 与图 6 展示了一些研究工作中使用的 BOO 分布图像及相关功能。

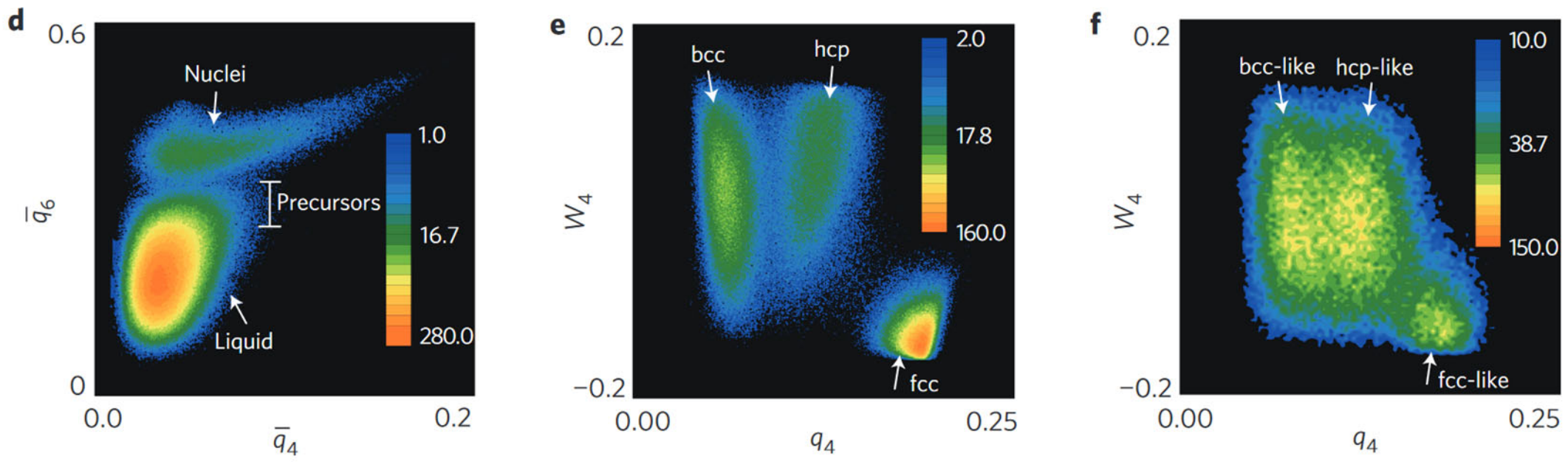

图 5：胶体成核实验中系统的 BOO 二维分布图像。图像摘取自参考文献[4]。图中的$W_l$为本文的$\widehat{W}_l$，图中的$\bar{q}_l$与本文的$\bar{Q}_l$计算方法有些许区别：其首先使用$S_{ij}$将原子分为固相与液相，而后$\bar{q}_l(i)$的计算过程中仅对与原子$i$相同相的近邻原子求和。在图一中，作者通过$\bar{q}_4\bar{q}_6$分布图像中聚集的两个峰值，将粒子分为液体与核；对液体种$\bar{q}_6$较大的部分单独分类为前驱体。图二为属于“核”的粒子的$\bar{q}_4W_l$分布。图三为属于前驱体的粒子的$\bar{q}_4W_l$分布。作者通过图二图三中不同的粒子聚集区域，将核粒子与前驱体粒子分别划分为如图的三种结构。颜色代表粒子计数（采用对数标度）。

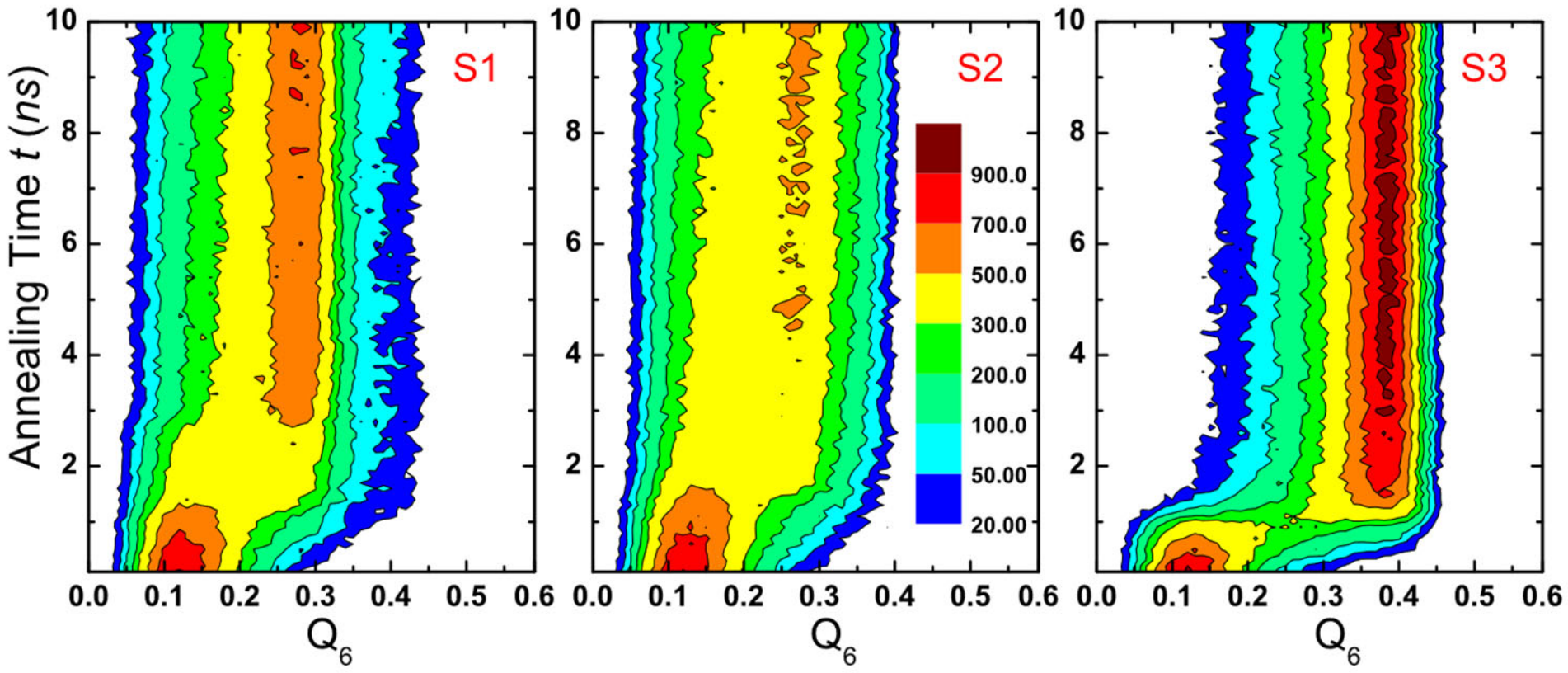

图 6：金属玻璃形成液体 ZrCu 的成核模拟中三组样品的 BOO 参量分布随时间演化分布图

像。颜色代表序参量为指定值的原子数量。图像摘取自参考文献[37]。图中所示的$Q_l$为本文中的$\bar{Q}_l$。可见，通过 BOO 参量分布随时间演化图像，可对体系何时发生相变以及相变过程中原子局域结构序参量的变化进行清晰而连续的描述。

## 2.4 平均首次通过时间方法

在通过原子的局域结构对原子所处的相进行分类，并且得到了体系不同时刻的成核反应坐标数值后（即最大固体团簇的粒子数$\hat{N}$）后，我们希望进一步得到成核问题中所关心的参量：体系的稳态成核速率与临界核大小。考虑到我们所做的第一性原理分子动力学模拟与（半）经验势分子动力学相比巨大的计算量，难以对体系进行充足的抽样以得到自由能曲线进而计算临界核大小。在此我们采用平均首次通过时间（Mean First-Passage Time，MFPT）方法低成本的对体系的成核率与临界核大小进行计算。MFPT 方法是 Wedekind 等人于 2007 年提出的用于处理对激活过程的分子动力学与布朗动力学模拟的方法[41]；Lundrigan 等人于 2009 年应用 MFPT 方法于 LJ 液体的成核问题[42]。

应用 MFPT 方法计算成核率与临界核大小是非常简洁的，首先我们在相同条件下运行多个过冷液体的成核的分子动力学模拟，对于每一个分子动力学轨迹，我们计算体系内最大固体团簇的原子数首次到达$n$的时间$\tau(n)$，这一时间即所谓的“首次通过时间”。 MFPT 方法证明可通过对多组分子动力学轨迹的$\tau(n)$取平均，而后按下式拟合，以得到$n_c$与$J$的估计值[41] [42]：

$$\tau(n)=\frac{1}{2JV}\{1+erf\,[c(n\text{-}n_c)\,]\} \qquad (2.4\text{-}1)$$

其中$erf$为误差函数，$c$为一拟合参量与 Zeldovich 因子有关[42]。

# 第3章 无轨道密度泛函分子动力学成核模拟

## 3.1 计算流程与计算参数

### 3.1.1 总研究思路

由二步成核理论[33,34]，前驱体和晶体核可以看作为两个独立的成核过程，在过冷液体中先出现前驱体，而后在前驱体内出现晶体核。在本工作我们对整个成核结晶过程的BOO序参量进行计算分析，同时对前驱体的成核的动力学参数进行了计算分析。

本工作主要分为三个阶段：1. 通过分子动力学进行两相法熔点测试；2. 进行批量分子动力学成核模拟，并自动在体系成核时停止模拟。3. 对体系进行后分析，采用键朝向序参量分布对体系局域结构进行分析，采用 MFPT 方法对体系动力学参数进行分析。对以上的每一阶段的分子动力学，作者皆采用无轨道密度泛函与 EAM 势能进行两次计算，并对两种计算方式的结果进行对比分析。以下对模拟中使用的软件、算法与计算参数进行介绍，不同阶段的更多模拟流程与细节在相应的小节中具体讨论。

### 3.1.2 无轨道密度泛函能量计算

本工作采用 ATLAS 实空间有限差分无轨道密度泛函软件[43]进行无轨道能量计算，动能密度泛选用 Wang-Teter 动能密度泛函[44]，实空间网格间选取$0.2Å^{-1}$，动能密度泛函中 Von-Weisaker 部分计算方式为在倒空间中计算，能量收敛阈值为$1.0\times10^{-5}eV/atom$，交换关联泛为 Perdew 与 Zunger 等人开发的 LDA 交换关联泛函（PZ81）[45]，赝势选取 Carter 等人基于 LDA 交换关联泛函开发的倒空间格式的铝的体相局域赝势（bulk local pseudopotentials，BLPS）[46,47]。

### 3.1.3 分子动力学计算

本工作通过 LAMMPS 分子动力学模拟软件[48]耦合 ATLAS 进行无轨道密度泛函分子动力学计算，其中 ATLAS 负责进行原子受力计算，LAMMPS 负责完成进一步

的分子动力学计算；通过 LAMMPS 调用 Mendelev 等人开发的铝的 EAM 势能[7]完成 EAM 势能分子动力学计算，注意这一 EAM 势能在开发时考虑了 Al 融化属性与液体结构，因此可以对铝过冷液体的成核动力学具有相对准确的描述，非常适合用来与我们的第一性原理成核模拟进行对比。除了在相应章节单独标注的特殊情况，本文使用的分子动力学步长为 2fs。本文使用 LAMMPS 实现的 NPT 与 NPH 系综模拟功能，其使用 Shinoda 等人发展的动力学方程[49]，其使用的控温器基于 Nose-Hoover Chain[50]，其使用的控压器由 Martyna 等人发展[51]。对于三组 NPH 系综熔点测试组控压器增加 Ptemp 参数，依次设为 950K,1000K 与 1100K，并使用 aniso 参数；对于 NPT 系综成核模拟控压器使用 iso 参数。经过测试，以上参数的选择皆在合理范围且对模拟无明显影响。降温细节将在后文中讨论。其余的参数皆使用默认或推荐值。

### 3.1.4 原子序参量计算

对于本文中两相法熔点计算，作者通过 OVITO 原子模型可视化与数据分析软件[52]进行多面体模板匹配方法（Polyhedral template matching）计算[53]，以快速准确的对各个原子按照局域结构划分为晶相结构或其他结构，进而对“体系是否完全液化或固化”进行快速而简便的分析。

对于本文中的成核模拟与其后的分析阶段，本工作则采用基于 BOO 参量的序参量进行分析处理。作者通过自己基于 PYSCAL 局域原子结构环境计算 Python 模块[54]开发的 GlassViewer 软件[55]对 BOO 参量进行计算。GlassViewer 是一个专注于金属玻璃体系与成核问题的序参量处理软件，其特点在于：1.可以对多分子动力学轨迹的序参量进行自动化并行处理；2.提供序参量的自动可视化；3.对一些 PYSCAL 所支持序参量计算提供优化，并提供更多的序参量计算支持，如键角分布（Bond Angle Distribution）、结构因子（Structure Factor）、Cargill-Spaepen 化学短程序（Cargill-Spaepen Chemical Short-range order）；3.提供首次通过时间（First-Passage Time）计算，以为成核问题提供支持；4.提供 BOO 参量随时间分布计算与 BOO 参量联合分布计算。其自行实现的功能与继承 PYSCAL 的功能列表可参见 GlassViewer 主页[55]。

### 3.1.5 原子构型可视化

在本工作中，均采用 OVITO 原子模型可视化与数据分析软件[52]对原子构型进行

可视化。

## 3.2 两相法熔点计算

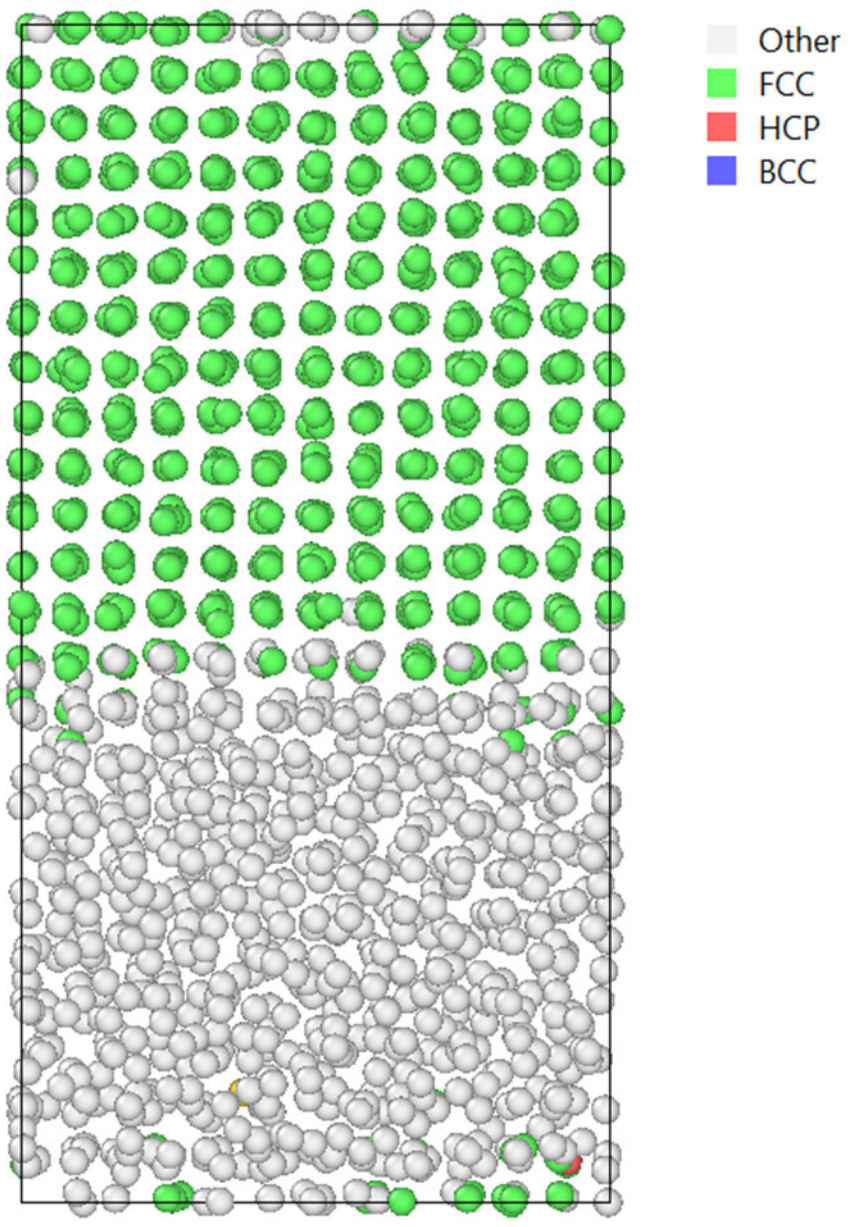

图 7：由 EAM-MD 为 OFDFT-MD 所准备的初始结构。

两相法即构造一个同时拥有液体、固体的超胞，通过合理调整初始参数，使在分子动力学中达到平衡时两相界面仍然存在，进而计算体系熔点。由于 KSDFT 计算量较大，难以对较多原子进行相对长时间的分子动力学模拟，传统通过 KSDFT 实现第一性原理分子动力学两相法熔点测试是非常困难的。OFDFT 中我们可以对 KSDFT 所难以计算的原子数规模进行分子动力学模拟，可以以两相法对体系熔点进行直观的测试。在两相法中我们必须非常仔细的准备体系的初始结构，为体系赋予合理的初始能量，以期体系能够实现两相共存的平衡，不会由于能量过高或过低导致超胞全部融化或全部凝固。我们共进行了三次 OFDFT-MD 模拟，每一组模拟流程如下：为了准备初始结构，我们首先准备一定体积$V_i$ $(i \in 1,2,3)$的 FCC 结构的 Al 的长方体超胞，由上下两个立方体超胞构成，每个立方体含有 864 个 Al 原子，共含有 1728 个 Al 原子。而后我们使用 EAM 势能分子动力学在 NVT 系综下固定上半晶胞的原子，让下半晶胞相继在 1800K、1500K 各模拟 15ps 以达到融化，然后固定下半晶胞，让上半

晶胞在 800K 模拟 15ps。这样做的目的是让两个相尽可能达到平衡的熔点附近的结构，i=1 时得到的初始结构如图 7 所示，$i = 2$与$i = 3$的初始结构与之类似。而后我们对超胞整体设定指定温度$T_i$ ($i \in 1,2,3$)的麦克斯韦分布速度，并对体系所有原子的总线速度、总角速度清零以防止质心漂移，而后运行 0 压下 NPH 系综下的 OFDFT-MD，使用 NPH 系综可以让体系在固定压力，固定焓值的情况下，自动将温度向熔点弛豫。经过 EAM-MD 测试，零压计算得到的熔点与标准大气压实验结果无明显区别，故使用零压进行模拟是可行的。

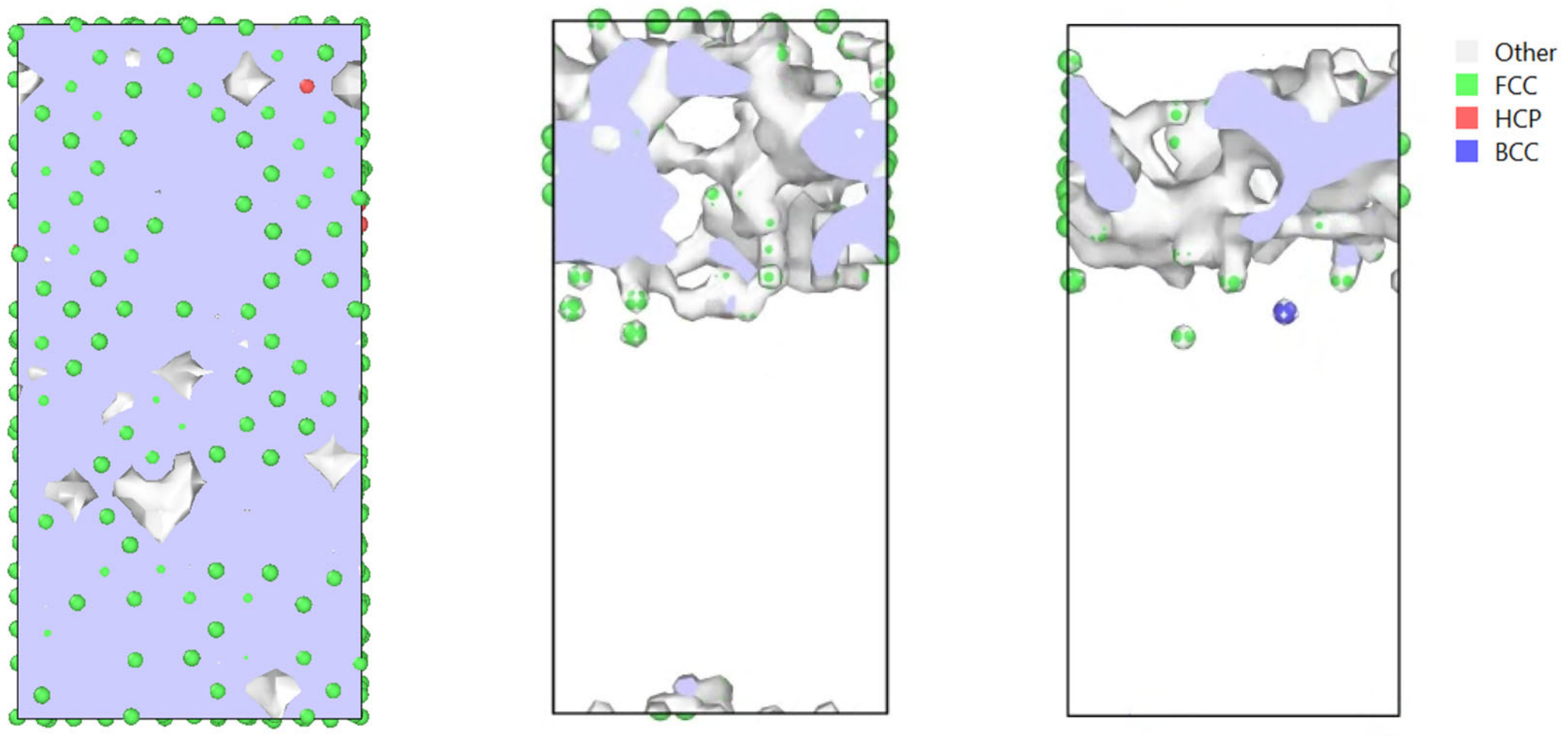

图 8：OFDFT-MD 两相法熔点测试原子构型可视化图像。从左至右分别为第一组、第二组与第三组临近模拟末尾处的原子构型。三组图像各自经过等比例缩放处理以考虑排版美观。由 Polyhedral template matching 方法对原子进行结构分析，将非晶相移处，不同晶相的原子被标记为不同颜色，如图例所示。作者通过 OVITO 软件采用高斯密度方法[56]为固体部分添加可视化表面以增强可分辨能力。可见，第一组全部凝固，而第二组与第三组保有两相界面。图中出现不规则可视化表面是由于熔点附近原子局域结构涨落较大以及可视化表面计算参数设置所引起。

首先我们进行第一组 OFDFT-MD 模拟，由于无法预知体系将会进一步融化或进一步凝固，$V_1$选取合理的体积值即可，$T_1$选为 850K，共模拟 35ps，结果发现体系在 15ps 左右全部凝固，如图 8 中第一组所示；我们进一步以第一组的平衡平均原子体积乘以原子数作为$V_2$，$V_3$，选取$T_2$=1000K，$T_3$=1100K 进行 OFDFT-MD 模拟，第二组模拟共进行 50ps，第三组模拟共进行 35ps。第二组第三组直至模拟完成，液固界面仍然保持存在，如图 8 中第二组与第三组所示。

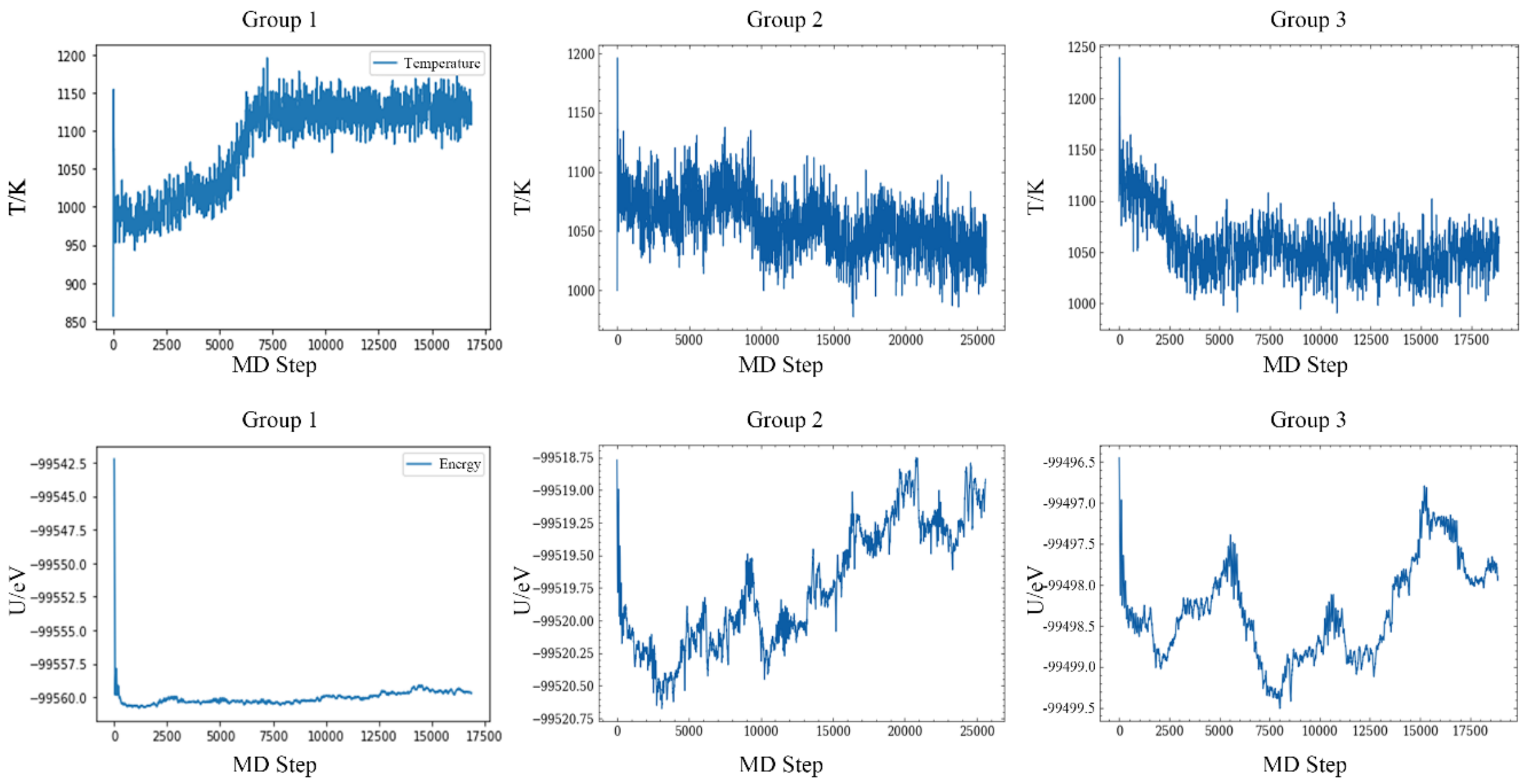

图 9：三组 OFDFT-MD 模拟体系的温度随时间变化曲线（上排）与总能随时间变化曲线（下排）。横坐标单位为模拟的步数，纵坐标单位为 K 与 eV。从左到右分别为第一组、第二组与第三组相对应的曲线图。注意各组曲线的横纵坐标范围并不相同，尤其注意第一组能量随时间变化曲线的纵坐标的变化范围较大。第一组初始巨大的能量下降可能由于初始体积设置与平衡态相差较大所致，这一现象在第二组第三组中使用修正的体积后得以改善。

图 9 中展示了三组 OFDFT-MD 模拟过程中的温度、总能随时间变化曲线。零压时 NPH 系综的总能量守恒，可见第一组由于初始体积设置的距离平衡态偏差较大，总能发生较大变化，但三组模拟的总能都在 1ps 内趋于稳定。同时，体系的温度可以描述体系的平均动能进而描述体系原子的总动能；由于体系总能守恒，因此温度可以进一步描述体系的总原子势能，进而描述相变的进展。从图中可见，总能的稳定时间远远小于相变发生的时间尺度。接下来，我们通过对不同体系的温度与总能进行分析，以分析体系熔点的取值。

表 1：三组 OFDFT-MD 模拟的总能近似值。可见第一组大于第二组大于第三组。

| | 第一组 | 第二组 | 第三组 |
|---|---|---|---|
| 总能近似值 | 99560 eV | 99520 eV | 99498 eV |

如表 1 所示，第一组第二组与第三组的总能依次递减，并在模拟中很快趋于稳定，通过温度曲线，我们可以发现，第一组的总动能逐渐升高，最终稳定在 1100K 至 1150K 之间；第二组的总动能略微下降，最终保持在 1050 左右；第三组的总动能下降的较第二组明显，最终也保持在 1050 左右。根据热力学知识可知，若不考虑界面

等其他效应，体系总能$E$应唯一确定系统的固体比例$n_s(E)$。当体系在零压熔点温度达到平衡时，若体系正好具有一半液相一半固相，则体系的总能、动能与势能是确定的，不妨设为$E_0$，$E_{K0}$，$E_{P0}$。在实际模拟中，我们通过对分别在熔点上、下温度处准备的一半固体一半液体的静止原子构型赋予不同温度的麦克斯韦速度分布，提供了三个势能相同（但不一定等于$E_{P0}$），动能与总能不同的远离平衡的原子构型。因此，当体系的总能正好等于$E_0$时，$n_s(E)$应等于 1/2；当体系的总能小于$E_0$时，多余的势能转换为不足的动能，体系进一步凝固，$n_s(E)$应小于 1/2；当体系的总能大于$E_0$时，多余的动能应转换未不足的势能，体系进一步融化，$n_s(E)$应大于 1/2。这正好与上述三组模拟中发现的现象相互吻合。

考虑到具有不同总能的第二组与第三组模拟的温度弛豫到非常相近的数值，且第二组与第三组平衡后的原子构型中都出现有两相的界面，我们不难推论，第二组与第三组的温度最终弛豫至体系的熔点，我们选取两组温度较为稳定的时间范围进行平均，第二组从 10000 步到 25000 进行平均，平均温度为 1045.60K，第三组从 5000 步到 11858 步进行平均，平均温度为 1047.16K。因此我们得到了体系的熔点：1046.38±0.78K，与 Al 在大气压下的实验熔点 933K 相比，偏差为 12.15%。可见基于第一性原理的 OFDFT-MD 模型可以相对较好的对 Al 的熔点进行预测。事实上，用于成核模拟的铁的 EAM 势能甚至可能将 1811K 的铁的实验熔点估计为 2400K。[5] OFDFT-MD 所预测的熔点与实验值的偏差主要源于所采用的 Wang-Teter 动能密度泛函。

注意读者可能会发现，在系统未完全凝固或融化的时候，系统的温度却发生上升与下降，而非稳定在某个熔点处，这是由于在体系达到平衡、属性达到稳定之前，只要系统的动能与势能的比例与其对应的平衡态不符，体系一定经历着非平衡的动能与势能相互转换的过程，只要这一转换过程正在进行，系统的动能便发生变化，因此温度也一定发生变化。尽管如此，这也无法解释第一组模拟在达到准平衡后的出现的“过热固体”现象，这可能是由于在凝固至熔点附近时，液相太少，涨落成一个半径小于“固体中的液体临界核” 的半径的液体核，进而在表面张力的作用下逐渐收缩消失。

接下来我们采用 EAM-MD 按照同样的流程进行模拟，得到熔点约为 926K，这一数据与原文献中熔点数据非常相近[7]，之所以其与大气压下实验值 933K 如此接近是

由于这一 EAM 势能本身基于熔点的数据所拟合。这一结果进一步说明了我们模拟的可靠性。同时在 EAM-MD 模拟中，也观察到了总能较低时出现的“过热固体”的奇异现象。

## 3.3 批量分子动力学成核模拟

在本工作中，我共计算 10 组 EAM-MD 成核模拟、8 组 OFDFT-MD 成核模拟。在成核模拟完成后，计算 OFDFT-MD 与 EAM-MD 两种方法对应的前驱体稳态成核速率$J_{pre}$与“$\bar{Q}_6$与$\widehat{W}_6$参量分布随时间变化的图像”。在成核模拟中我们选取 10K/ps 的降温速率，将体系从 1140K 降温至 300K 后使之恒温。与 933K 的熔点相比，300K 的体系过冷度约 633K。与此相近的降温速率与过冷度在之前的 Al 的（半）经验势能成核模拟工作中曾被使用[57]，其证明了在这一参数下的成核处于计算机模拟可以计算的范围之内。每组模拟在降温前被赋予一个随机的 1140K 的麦克斯韦速度分布以保证不同组的随机性。为了避免尺寸依赖性对模拟的影响，经过收敛性测试，体系原子数选为 864 原子，尺寸依赖收敛性测试的具体细节将在本小节末给出。

成核模拟的分子动力学模拟与一般的固定时长的分子动力学所不同，体系成核所需时间是随机的，在未成核前应让体系持续保持模拟，在模拟中每隔一段时间对体系的序参量进行计算，进而判断体系是否成核以决定是否停止模拟。在本工作中使用自动化设计自动进行上述判断流程，对于 EAM-MD 模拟每 12000 步进行一次判断，对于 OFDFT-MD 每 2000 步进行一次判断。在每次成核判断中，首先使用 SANN 方法[58]（solid-angle based nearest-neighbor algorithm）无参数的为每个原子求得合适的截至半径，从而求得每个原子的近邻原子，而后我们对体系原子的$\bar{Q}_6$进行计算。并筛选出$\bar{Q}_6 > 0.3$的原子，在后文中我们可以看到$\bar{Q}_6$大于这一阈值的原子可以被认为是二步成核中的“前驱体与晶体”原子。若体系的“前驱体与晶体”原子占比小于 95%则持续进行计算。若体系的“前驱体与晶体”原子占比大于 95%时，则认为系统已经成核，此时让 EAM-MD 再运行 12000 步，OFDFT-MD 再运行 2000 步，以让体系更充分的相变，而后自动停止模拟。

在模拟结束后，为了计算前驱体的成核率$J_{pre}$与前驱体临界核原子数$n_{c,pre}$，我们首先对分子动力学轨迹每 200fs 采样一次，使用 SANN 方法为每一原子选择截至半径以计算每一原子的近邻原子，进而计算每个原子的$\bar{Q}_6$并筛选出$\bar{Q}_6 > 0.3$的“前驱体与晶体”原子；之后对被筛选出的“前驱体与晶体”原子按照 SANN 方法得到的截至半径进行团簇划分；按照体系的“前驱体与晶体”原子的最大团簇的原子数$\hat{N}_{pre}$，我们可以计算一个分子动力学轨迹的“首次通过时间”。通过对 10 组 EAM-MD 分子动力学轨迹的首次通过时间进行平均，进而得到平均首次通过时间曲线以拟合前驱体的成核率$J_{pre}$与前驱体临界核原子数$n_{c,pre}$。

对于 BOO 随时间分布图像的计算，我们首先对分子动力学轨迹每 2ps 采样一次，对每个时间采样，通过程序自动拟合出对关联函数中第一个峰与第二个峰之间的极小值位置，以此作为当前时间采样内所有原子的近邻截至半径以计算每个原子的近邻原子。而后计算每个原子的$\bar{Q}_6$与$\hat{W}_6$，以求得当前时间采样的$\bar{Q}_6$与$\hat{W}_6$分布。进而我们可以求得$\bar{Q}_6$与$\hat{W}_6$分布随时间变化的图像。之所以此处对每一时间采样以拟合的方式求近邻原子截至半径，是想为同一时间采样内的全部原子设定相同的截至半径（而非 SANN 方法中每一个原子有单独的截至半径）。由于近邻原子数的选择会对$\bar{Q}_6$的值造成影响，通过为一个时间采样内的全部原子设定统一的截止半径，可以使得同一时间采样内相对紧密的与相对松散的不同的局域原子结构拥有区分度更加明显的$\bar{Q}_6$，从而使我们能够通过$\bar{Q}_6$的分布更好的对系统的的局域结构进行表征。之所以对每一时间采样单独拟合，而非为所有时间采样设立同一个近邻截至半径，是考虑降温与相变过程中系统体积的变化会造成对关联函数第一壳层的位置发生变化，对每一时间采样单独拟合可以防止局域结构相同的原子由于在降温中密度发生变化而使其$\bar{Q}_6$值发生偏移。

这里应指出，OFDFT-MD 使用与 EAM-MD 不同的模拟起始结构。OFDFT-MD 的起始结构使用 OFDFT-MD 在 1140K 弛豫至少 10ps 得到的结构。EAM-MD 的起始结构则使用 EAM-MD 在 1140K 弛豫至少 10ps 得到的结构。准备起始结构时，弛豫至少 10ps 是为了保证系统融化的更加完全。之所以严谨的准备不同的起始结构，是由于温度为 1140K 时，不同模拟方法有其不同的平衡构型，同时我们使用的降温速率下的降温成核是非常快的非平衡过程，如使用非某方法对应的起始平衡构型，无法保

证系统在模拟过程中被弛豫至这个方法所对应的成核动力学路径上，从而对模拟结果造成影响。因此我们应保证每个方法使用其自身对应的起始平衡结构进行非平衡降温以保证模拟的可靠性。

注意到我们使用“前驱体与晶体”原子团簇的平均通过时间曲线对前驱体成核动力学参数进行拟合，这是由于二步成核现象与二步成核理论中，前驱体先于晶体出现，晶体在前驱体中出现，故对“前驱体与晶体原子”团簇的统计可以反应前驱体成核的情况。

对于过长时间仍未实现“前驱体与晶体”原子占比 95%的体系，由人为确定其“前驱体与晶体”原子占有相对较多比例后手动关停。“前驱体与晶体”原子占有相对较多比例，但无法达到 95%的现象可能是由于“不同团簇晶界的错位等原因使得体系难以在序参量的意义下相变完全”。

在尺寸依赖性收敛性测试中，我对不同原子数（从 108 原子至 2916 原子）的系统使用 EAM-MD 计算$J_{pre}$，发现当体系原子数增加至 864 原子时 $J_{pre}$收敛。在尺寸依赖性测试中使用的参数与正式模拟使用的参数略有不同：系统不经过降温而被直接施加以 570K 的控温器，MD 步长选为 2.5fs，对 MD 轨迹以 250fs 为时间间隔进行抽样。同样使用 SANN 方法寻找近邻原子，筛选出$\bar{Q}_6$大于 0.3 的原子以计算 MFPT 与$J_{pre}$。

## 3.4 局域结构与动力学参数分析

我们选取上文得到的 864 原子的 EAM-MD 与 OFDFT-MD 的模拟数据进行对比分析。我们首先对“前驱体与固体”原子的 MFPT 曲线进行分析。如图 10 中第一排两图所示，两种方法的 MFPF 曲线都在 5 个原子数附近存在一个明显的“折点”，小于五个原子数时，MFPT 上升的非常陡峭，大于五个原子数后，MFPT 上升的相对缓慢。这意味着，平均而言，在体系中最大的团簇的原子数由 0 演化至 100 的过程中，以很慢的速率在$t_1$演化至 5 个原子数，而后以较快的速率在$t_2$演化至 100 个原子数。由图像进行简单的估测与估算，OFDFT-MD 的 MFPT 曲线中$t_2/t_1$约为 2.1；在 EAM-MD 中约为 2.25。可见，体系中的核团簇仅仅以到达 5 个原子约两倍的时间就到达了 100 个原子！这充分说明了体系在 5 个原子左右完成成核，而后团簇进入了较快的生长阶段。考虑到 MFPT 曲线拟合中需要对误差函数形的曲线进行拟合，且误差函数的中点为临界核团簇原子数，为了更好的对成核阶段进行拟合，我们取 10 个原子以下的 MFPT 曲线进行拟合。从图 10 中第二排的拟合数据来看，其拟合效果远远好于图 10 中第一排用 100 个原子内的 MFPT 曲线进行拟合的效果：其更加贴合小原子数数据；更加体现了误差函数的形状；误差函数的中点也与我们分析的 5 个原子左右的核团簇原子数相吻合。

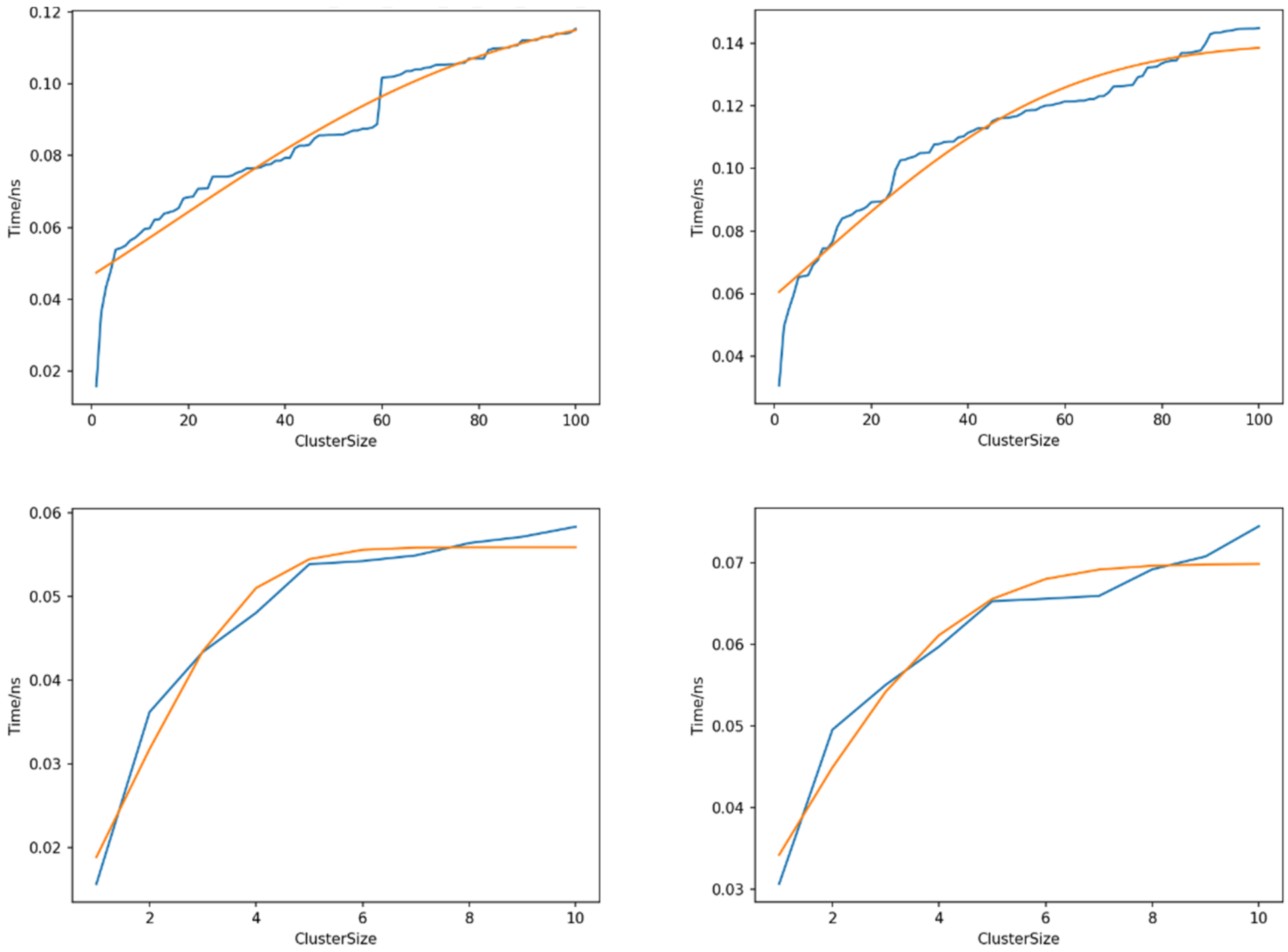

**图 10：OFDFT-MD 与 EAM-MD 成核模拟的 MFPT 曲线。第一行为 100 个原子数以内的 MFPT，第二行图像为 10 个原子数以内的 MFPT，每行左侧为 OFDFT-MD 图像，右侧为 EAM-MD 图像。图中蓝色线为模拟数据，黄色线为 MFPT 拟合曲线。可见第二排图像对成核对应的小原子数区域（如正文所分析）的拟合效果好于第一排。**

而后我们按照 MFPT 对前驱体的成和速率与临界核原子数进行计算。在计算中，为了得到相对合理的体系的体积，我们首先对两种方法的 MFPT 曲线按照（2.4-1）式拟合以得到两种方法对应的前驱体临界核大小与两种方法对应的前驱体临界核的出现所需的平均演化时间$\bar{t}_{EAM}$和$\bar{t}_{OFDFT}$（即拟合函数对称中心对应的纵坐标与横坐标）。而后计算 EAM-MD 全部模拟组在$\bar{t}_{EAM}$时的平均体积，以及 OFDFT-MD 在$\bar{t}_{OFDFT}$

时的平均体积。进而按照（2.4-1）式对 MFPT 曲线拟合即可得到两种方法所对应的前驱体稳态成核速率。两种方法所对应的前驱体稳态成核速率与前驱体临界核大小数据如表 2 所示。可见两种方法模拟得到的前驱体成核速率与前驱体临界核原子数不仅在数量级上相一致，在数量上也非常相近。两种方法预测的$J_{pre}$的偏差仅为 24%。

表 2：OFDFT-MD 与 EAM-MD 模拟结果通过 MFPT 方法拟合的前驱体成核速率$J_{pre}$与前驱体临界核原子数$n_{c,pre}$。

| Method | $J_{pre}(\mathrm{m^{-3}s^{-1}})$ | $n_{c,pre}$ |
|---|---|---|
| OFDFT-MD | $1.18\times10^{36}$ | 1.71 |
| EAM-MD | $8.98\times10^{35}$ | 1.06 |

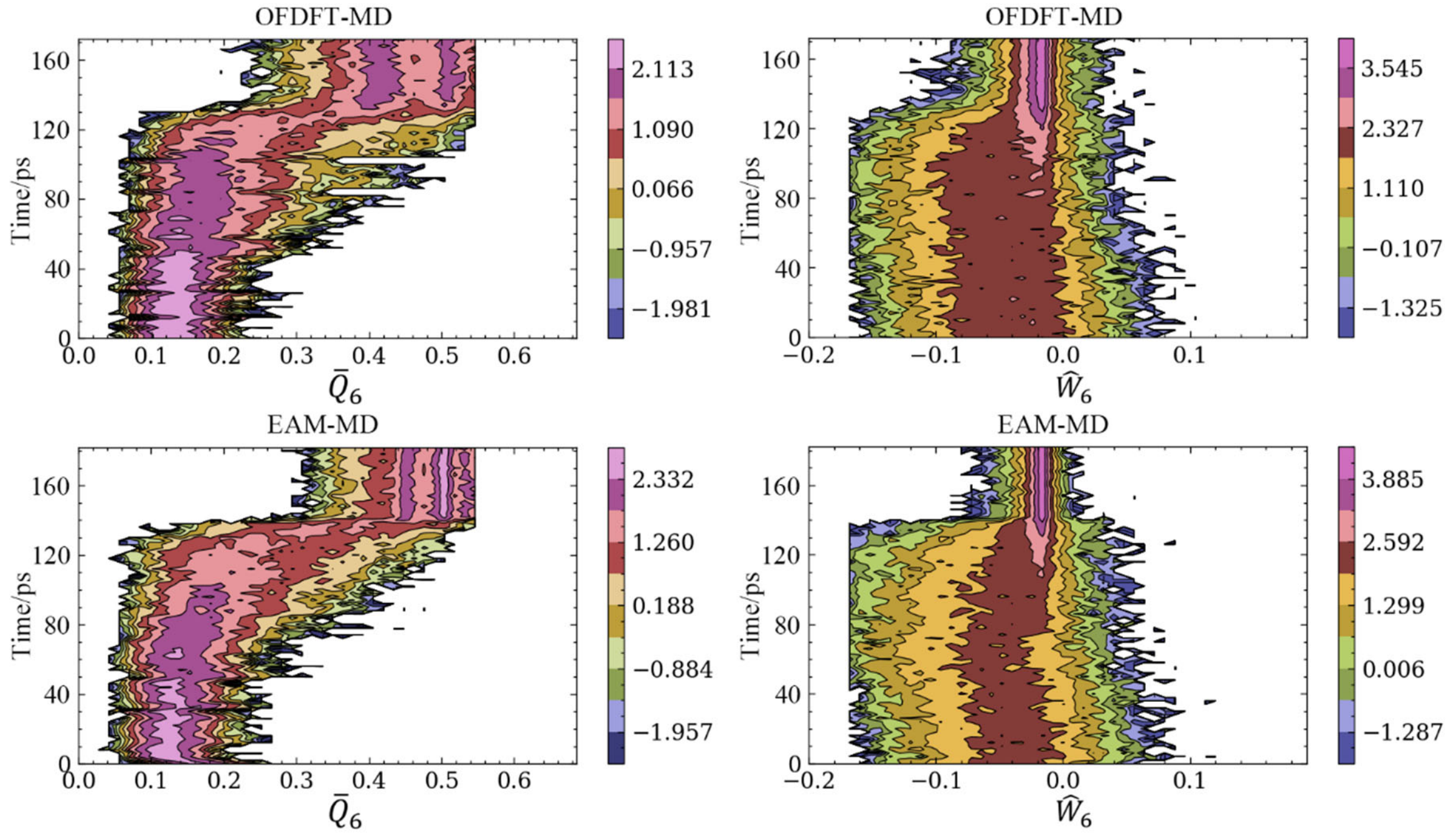

图 11：OFDFT-MD 与 EAM-MD 成核时间相近的两组模拟的 BOO 参量对比（对数标度）。

最后我们对两种方法得到的 BOO 参量随时间分布进行对比，由于同一种方法不同模拟组的成核时间不同，我们选取两种方法成核时间相近的两组模拟进行对比。由图 11 可见，两种方法得到的 BOO 参量随时间分布图像大体相同：由$\bar{Q}_6$随时间分布图像可见，系统在 50ps 处出现向右扩展的$\bar{Q}_6$分布，并逐渐出现$\bar{Q}_6$大于 0.3 的原子局域结构，而后在 120-160ps 中间完成快速的晶化成核，体系的序参量分布发生剧烈的变化。这一物理图景正对应二步成核给出的物理图景：$\bar{Q}_6$大于 0.3 的原子的出现对应着前驱体的出现，随着前驱体原子的增多，在前驱体中出现晶核。从这个分布图像上

也可看出我们以 0.3 作为前驱体的$\bar{Q}_6$阈值是合理的。但同时我们也观察到，两种方法的$\bar{Q}_6$图像也有些许区别，如晶化后的峰值分布的位置。对于$\widehat{W}_6$图像，两种方法表现的非常相似。

两种方法的结果的相似性表面上看是平庸的，起始不然，我们利用的是两种较为不同的物理模型。我们采用的 EAM 势能是基于相对合理的原子相互作用假设，依照熔点附近材料的实际性质拟合而成的微观模型。而 OFDFT 是基于第一性原理计算方法，不需要经验拟合参数，其计算精度主要由所采用的动能密度泛函决定。两个出发点较为不同的方法却预测出相近的结果，这同时验证着 OFDFT 与所采用的 EAM 势能对铝过冷液体成核过程中的势能面的建模是相对准确的。我们的结果验证了 OFDFT-MD 用于过冷液体的成核研究中的可行性。

# 第4章 总结与展望

在本论文中，作者首先对一阶相变与成核现象的性质、成核研究的产业价值、成核问题的研究困难与成核问题的研究缺口进行了介绍；而后对经典成核理论、计算机模拟方法(DFT/OFDFT、EAM 与 MD 理论)、二步/多步成核、键朝向序参量与 MFPT 方法等理论进行了综述；最后介绍了作者使用 OFDFT-MD 与（熔点附近拟合的）EAM-MD 对过冷液铝成核的模拟研究，并得出以下结论：

1. 通过 1728 原子的两相法测试，零压下 OFDFT-MD 模拟得到的铝的熔点为 1046.38±0.78K，与大气压下实验熔点 933K 相比偏差为 12.15%。
2. 以 10K/ps 的降温速率降温至 300K，864 原子的 8 组 OFDFT-MD 模拟与 10 组 EAM-MD 模拟对过冷液铝的前驱体的成核动力学参数给出的预测不仅处于同一个数量级、数值结果亦十分接近，两种方法预测的前驱体的稳态成核速率偏差仅为 24%；
3. 选取两种方法成核时间相近的两组模拟结果，其表现出大体相同的$\bar{Q}_6$随时间分布图像，其中观察到晶化后$\bar{Q}_6$峰值的分布的位置略有区别；除此以外，这两组模拟结果表现出十分相似的$\widehat{W}_6$随时间分布图像。

本工作一方面论证了 OFDFT-MD 用于过冷液体成核研究的可行性，由于第一性原理模拟引入了电子自由度，其为过冷液体成核过程中电子结构的研究提供了可能，也为对过冷液体成核过程的电磁场调控的研究提供了切入点。另一方面，在本工作中，分别基于“第一性原理电子结构计算”与“采用熔点附近材料性质拟合的半经验势能”的两种构造方式差别较大的计算方法，却给出如此相似的成核的动力学结果，其背后的物理内涵值得深思：金属过冷液体的前驱体的成核速率是否并不依赖于具体的微观模型，而仅依赖于不同微观模型在某一尺度下的某种特征（正如理想气体模型的体积依赖于压强与温度）？如果的确是这样，这个独特的物理参量是什么？我们是否可以依据它构建出一个类似经典成核理论，但具有相对准确的预测能力的模型？同时，在本模拟工作中仍有未被探索的具有价值的研究问题：如“对于晶相的成核过程（非前驱体的），两种方法是否也会表现出相似的动力学参数”、“可否通过 BOO 参量的联合分布进一步发现两种方法成核路径的细微差别”等。最后需要指出的是，OFDFT 由于使用动能密度泛函代替 KS 无相互作用动能，与 KSDFT 相比能量计算

误差可能较大，这类误差是否会较大的影响对成核的模拟亦待考究。

# 参考文献

[1] So Much More to Know...[J]. Science, 2005, 309(5731): 78-102.
[2] Allen R J, Valeriani C, Ten Wolde P R. Forward flux sampling for rare event simulations[J]. Journal of Physics: Condensed Matter, 2009, 21(46): 463102.
[3] Sanz E, Vega C, Espinosa J R, et al. Homogeneous Ice Nucleation at Moderate Supercooling from Molecular Simulation[J]. Journal of the American Chemical Society, 2013, 135(40): 15008-15017.
[4] Tan P, Xu N, Xu L. Visualizing kinetic pathways of homogeneous nucleation in colloidal crystallization[J]. Nature Physics, 2014, 10(1): 73-79.
[5] Shibuta Y, Sakane S, Miyoshi E, et al. Heterogeneity in homogeneous nucleation from billion-atom molecular dynamics simulation of solidification of pure metal[J]. Nature communications, 2017, 8(1): 1-9.
[6] Debela T T, Wang X D, Cao Q P, et al. Nucleation driven by orientational order in supercooled niobium as seen via ab initio molecular dynamics[J]. Physical Review B, 2014, 89(10): 104205.
[7] Mendelev M I, Kramer M J, Becker C A, et al. Analysis of semi-empirical interatomic potentials appropriate for simulation of crystalline and liquid Al and Cu[J]. Philosophical Magazine, 2008, 88(12): 1723-1750.
[8] 米文慧. 基于无轨道密度泛函理论的 OEPP 赝势与实空间 ATLAS 计算软件[J]. 吉林大学博士论文.
[9] Hayes R L, Fago M, Ortiz M, et al. Prediction of dislocation nucleation during nanoindentation by the orbital-free density functional theory local quasi-continuum method[J]. Multiscale Modeling & Simulation, 2005, 4(2): 359-389.
[10] Hayes R L, Ho G, Ortiz M, et al. Prediction of dislocation nucleation during nanoindentation of Al3Mg by the orbital-free density functional theory local quasicontinuum method[J]. Philosophical Magazine, 2006, 86(16): 2343-2358.
[11] Zeldovich Y B. On the theory of new phase formation: cavitation[J]. Acta Physicochem., USSR, 1943, 18: 1.
[12] Volmer M. Kinetik der Phasenbildung, Steinkopff, Dresden u[J]. Stuttgart, Germany, 1939.
[13] Becker R, Döring W. Kinetische behandlung der keimbildung in übersättigten dämpfen[J]. Annalen der physik, 1935, 416(8): 719-752.
[14] Kashchiev D. On the relation between nucleation work, nucleus size, and nucleation rate[J]. The Journal of Chemical Physics, 1982, 76(10): 5098-5102.
[15] Kalikmanov V I. Nucleation Theory: Vol. 860[M]. Dordrecht: Springer Netherlands, 2013.
[16] Turnbull D, Fisher J C. Rate of nucleation in condensed systems[J]. The Journal of chemical physics, 1949, 17(1): 71-73.
[17] Sear R P. Nucleation: theory and applications to protein solutions and colloidal suspensions[J]. Journal of Physics: Condensed Matter, 2007, 19(3): 033101.
[18] Prestipino S, Laio A, Tosatti E. Systematic Improvement of Classical Nucleation

Theory[J]. Physical Review Letters, 2012, 108(22): 225701.
[19] Wölk J, Strey R. Homogeneous nucleation of H2O and D2O in comparison: the isotope effect[J]. The Journal of Physical Chemistry B, 2001, 105(47): 11683-11701.
[20] Labetski D G, Holten V, Van Dongen M E H. Comment on" The nucleation behavior of supercooled water vapor in helium[J]. The Journal of chemical physics, 2004, 120: 6314.
[21] Hohenberg P, Kohn W. Inhomogeneous electron gas[J]. Physical review, 1964, 136(3B): B864.
[22] Kohn W, Sham L J. Self-consistent equations including exchange and correlation effects[J]. Physical review, 1965, 140(4A): A1133.
[23] Daw M S, Baskes M I. Semiempirical, quantum mechanical calculation of hydrogen embrittlement in metals[J]. Physical review letters, 1983, 50(17): 1285.
[24] Daw M S, Baskes M I. Embedded-atom method: Derivation and application to impurities, surfaces, and other defects in metals[J]. Physical Review B, 1984, 29(12): 6443.
[25] Finnis M W, Sinclair J E. A simple empirical N-body potential for transition metals[J]. Philosophical Magazine A, 1984, 50(1): 45-55.
[26] Parrinello M, Rahman A. Crystal structure and pair potentials: A molecular-dynamics study[J]. Physical Review Letters, 1980, 45(14): 1196.
[27] Parrinello M, Rahman A. Polymorphic transitions in single crystals: A new molecular dynamics method[J]. Journal of Applied Physics, 1981, 52(12): 7182-7190.
[28] 陈敏伯. 计算化学：从理论化学到分子模拟[M]. 科学出版社, 2009.
[29] Hoover W G. Canonical dynamics: Equilibrium phase-space distributions[J]. Physical review A, 1985, 31(3): 1695.
[30] Nosé S. A unified formulation of the constant temperature molecular dynamics methods[J]. The Journal of chemical physics, 1984, 81(1): 511-519.
[31] Basconi J E, Shirts M R. Effects of Temperature Control Algorithms on Transport Properties and Kinetics in Molecular Dynamics Simulations[J]. Journal of Chemical Theory and Computation, 2013, 9(7): 2887-2899.
[32] Chung S Y, Kim Y M, Kim J G, et al. Multiphase transformation and Ostwald's rule of stages during crystallization of a metal phosphate[J]. Nature physics, 2009, 5(1): 68-73.
[33] Iwamatsu M. Free-energy landscape of nucleation with an intermediate metastable phase studied using capillarity approximation[J]. The Journal of chemical physics, 2011, 134(16): 164508.
[34] Eaton D, Saika-Voivod I, Bowles R K, et al. Free energy surface of two-step nucleation[J]. The Journal of chemical physics, 2021, 154(23): 234507.
[35] Liu X, Chee S W, Raj S, et al. Three-step nucleation of metal–organic framework nanocrystals[J]. Proceedings of the National Academy of Sciences of the United States of America, 2021, 118(10): e2008880118.
[36] Takahashi K Z, Aoyagi T, Fukuda J ichi. Multistep nucleation of anisotropic molecules[J]. Nature Communications, 2021, 12(1): 5278.
[37] Jiang S Q, Wu Z W, Li M Z. Effect of local structures on crystallization in deeply

undercooled metallic glass-forming liquids[J]. The Journal of chemical physics, 2016, 144(15): 154502.
[38] Steinhardt P J, Nelson D R, Ronchetti M. Bond-orientational order in liquids and glasses[J]. Physical Review B, 1983, 28(2): 784.
[39] Lechner W, Dellago C. Accurate determination of crystal structures based on averaged local bond order parameters[J]. The Journal of chemical physics, 2008, 129(11): 114707.
[40] Dr. Holm C, Prof. Dr. Kremer K. Advanced Computer Simulation: Approaches for Soft Matter Sciences I: Vol. 173[M]. Berlin, Heidelberg: Springer Berlin Heidelberg, 2005.
[41] Wedekind J, Strey R, Reguera D. New method to analyze simulations of activated processes[J]. The Journal of chemical physics, 2007, 126(13): 134103.
[42] Lundrigan S E, Saika-Voivod I. Test of classical nucleation theory and mean first-passage time formalism on crystallization in the Lennard-Jones liquid[J]. The Journal of chemical physics, 2009, 131(10): 104503.
[43] Mi W, Shao X, Su C, et al. ATLAS: A real-space finite-difference implementation of orbital-free density functional theory[J]. Computer Physics Communications, 2016, 200: 87-95.
[44] Wang L W, Teter M P. Kinetic-energy functional of the electron density[J]. Physical Review B, 1992, 45(23): 13196.
[45] Perdew J P, Zunger A. Self-interaction correction to density-functional approximations for many-electron systems[J]. Physical Review B, 1981, 23(10): 5048.
[46] Zhou B, Wang Y A, Carter E A. Transferable local pseudopotentials derived via inversion of the Kohn-Sham equations in a bulk environment[J]. Physical Review B, 2004, 69(12): 125109.
[47] Huang C, Carter E A. Transferable local pseudopotentials for magnesium, aluminum and silicon[J]. Physical Chemistry Chemical Physics, 2008, 10(47): 7109-7120.
[48] Thompson A P, Aktulga H M, Berger R, et al. LAMMPS - a flexible simulation tool for particle-based materials modeling at the atomic, meso, and continuum scales[J]. Computer Physics Communications, 2022, 271: 108171.
[49] Shinoda W, Shiga M, Mikami M. Rapid estimation of elastic constants by molecular dynamics simulation under constant stress[J]. Physical Review B, 2004, 69(13): 134103.
[50] Martyna G J, Klein M L, Tuckerman M. Nosé–Hoover chains: The canonical ensemble via continuous dynamics[J]. The Journal of chemical physics, 1992, 97(4): 2635-2643.
[51] Martyna G J, Tobias D J, Klein M L. Constant pressure molecular dynamics algorithms[J]. The Journal of chemical physics, 1994, 101(5): 4177-4189.
[52] Stukowski A. Visualization and analysis of atomistic simulation data with OVITO-the Open Visualization Tool[J]. MODELLING AND SIMULATION IN MATERIALS SCIENCE AND ENGINEERING, 2010, 18(1).
[53] Larsen P M, Schmidt S, Schiøtz J. Robust structural identification via polyhedral template matching[J]. Modelling and Simulation in Materials Science and Engineering, 2016, 24(5): 055007.

[54] Menon S, Leines G D, Rogal J. pyscal: A python module for structural analysis of atomic environments[J]. Journal of Open Source Software, 2019, 4(43): 1824.

[55] Bai Z. GlassViewer[CP/OL]. https://github.com/bgbaizh/GlassViewer.

[56] Krone M, Stone J E, Ertl T, et al. Fast Visualization of Gaussian Density Surfaces for Molecular Dynamics and Particle System Trajectories.[J]. EuroVis (Short Papers), 2012, 10: 067-071.

[57] Papanikolaou M, Salonitis K, Jolly M, et al. Large-scale molecular dynamics simulations of homogeneous nucleation of pure aluminium[J]. Metals, 2019, 9(11): 1217.

[58] van Meel J A, Filion L, Valeriani C, et al. A parameter-free, solid-angle based, nearest-neighbor algorithm[J]. The Journal of chemical physics, 2012, 136(23): 234107.